# Using Generative Models to Produce Realistic Populations of UK Windstorms

Yee Chun Tsoi,[a] Kieran M. R. Hunt,[a,b] Len Shaffrey,[a,b] Atta Badii,[c] Richard Dixon,[a] Ludovico Nicotina[d]

[a] *Department of Meteorology, University of Reading, United Kingdom*

[b] *National Centre for Atmospheric Science, University of Reading, United Kingdom*

[c] *Department of Computer Science, University of Reading, United Kingdom*

[d] *Catastrophe Research, Inigo Insurance, London, United Kingdom*

*Corresponding author*: Kieran M. R. Hunt, k.m.r.hunt@reading.ac.uk

ABSTRACT

This study evaluates the potential of generative models, trained on historical ERA5 reanalysis data, for simulating windstorms over the UK. Four generative models, including a standard GAN, a WGAN-GP, a U-net diffusion model, and a diffusion-GAN were assessed based on their ability to replicate spatial and statistical characteristics of windstorms. Different models have distinct strengths and limitations. The standard GAN displayed broader variability and limited alignment on the PCA dimensions. The WGAN-GP had a more balanced performance but occasionally misrepresented extreme events. The U-net diffusion model produced high-quality spatial patterns but consistently underestimated windstorm intensities. The diffusion-GAN performed better than the other models in general but overestimated extremes. An ensemble approach combining the strengths of these models could potentially improve their overall reliability. This study provides a foundation for such generative models in meteorological research and could potentially be applied in windstorm analysis and risk assessment.

SIGNIFICANCE STATEMENT

Understanding windstorm patterns is crucial for mitigating their impacts to society and infrastructure, yet limited observations of extremes make it difficult to model their impacts accurately. This study explores generative models as tools for simulating windstorm characteristics over the UK. While these models demonstrate the ability to replicate key windstorm patterns, their stability and ability in capturing extremes remains challenging. This study demonstrates a novel approach in utilizing generative models to produce synthetic meteorological dataset, which can produce a large set of extreme windstorms, leading to an improved understanding of risk for the insurance and risk management industries.

## 1. Introduction

Windstorms are among the most impactful natural hazards in the United Kingdom, causing widespread disruption to society, infrastructure, and the economy. The geographical location of the country exposes it to windstorms, which frequently develop over the North Atlantic and especially affect the country during winter (Adam et al., 2016). Major historical events such as the Great Storm of 1987 and Cyclone Daria (Burns' Day storm of 1990) have demonstrated the potential severity of such events, resulting in billions of pounds in damage and significant loss of life (Cusack, 2023). Accurate catalogues of windstorms and their losses are essential for risk assessment, particularly in the insurance and re-insurance industries, where understanding the likelihood and severity of extreme events is crucial for financial planning and pricing risk.

However, modelling risks associated with extreme windstorms remain challenging due to their rarity. While moderate windstorms occur frequently, extreme events are rare, resulting in limited observational data. This scarcity of data poses a significant challenge for the development of catastrophe models, which are used to estimate the likelihood and impact of potential extreme events (Clark, 2002). The current approach is to apply regional climate models (RCMs) to simulate potential extreme events at high spatial resolution (Haylock, 2011). While these provide detailed and physically consistent samples of windstorms, they are time-consuming and scientifically challenging, making them less practical for the large-scale generation of event wind fields required to build extensive hazard datasets (Ashfaq et al., 2016).

Recent advancements in machine learning have introduced new methodologies for overcoming the limitations of traditional approaches. Generative models, such as generative adversarial networks (GANs) and denoising diffusion models are a class of unsupervised learning algorithms that generate new data samples that closely resemble the training data on a sample-by-sample basis. Once trained, they require much less computational resources to produce massive samples and have shown promise in replicating physical fields and generating realistic synthetic datasets (de Melo et al., 2022). By training on reanalysis data, these models aim to produce high-quality wind fields that capture the spatial variability and intensity of real-world windstorms. The generated data samples can therefore offer a solution to supplement traditional reanalysis datasets, which address the challenges posed by limited historical data, especially for rare but high-impact events.

The application of generative models, particularly GANs and denoising diffusion models, in meteorology has gained attention in recent years. GANs, initially introduced by Goodfellow et al. (2014), consist of two distinct neural networks, the generator and the discriminator. They are trained in a simultaneous adversarial process, with the generator trying to generate data samples that fool the discriminator, and the discriminator aiming to correctly classify them as real or fake. On the other hand, denoising diffusion models learn to reverse a gradual noise addition process and once trained, can generate high-quality data samples from pure noise (Ho et al., 2020).

One of the most common applications of generative models has been in precipitation nowcasting. Ravuri et al. (2021) introduced a deep generative model capable of producing radar-based rainfall forecasts with improved accuracy over traditional methods. In another study, Wang et al. (2023) proposed a task-segmented GAN architecture (TS-RainGAN), which successfully captured the spatiotemporal dynamics of heavy rainfall systems and improved skill scores and image quality for short lead times. Similarly, Asperti et al. (2023) applied a generative diffusion model that demonstrated improvements in rainfall intensity predictions compared to other U-net based models.

In addition to nowcasting, generative models have been widely applied in statistical downscaling of climate model output, a process that increases the spatial resolution of weather variables. Leinonen et al. (2020) introduced a stochastic super-resolution GAN that improves the spatial resolution of time-evolving atmospheric fields, while still capturing fine-scale variability. Similarly, Miralles et al. (2022) applied a GAN-based approach to downscale historical wind fields over Switzerland, demonstrating its ability to replicate fine-scale wind patterns and improved wind predictions over both flat and complex terrains.

Recent research has also explored the use of generative models for weather and climate simulation. For instance, Brochet et al. (2023) developed a residual WGAN model, which demonstrated excellent performance in generating synthetic multivariate atmospheric fields from the AROME Ensemble Prediction System (AROME-EPS). Similarly, Besombes et al. (2021) explored the potential of GANs to produce realistic climate data that could capture the multivariate distribution of weather variables in the Planet Simulator (PlaSim) while maintaining geostrophic balance.

Despite the growing interest in using generative models in meteorology, their application in windstorm simulation has been very limited. Although previous research has successfully demonstrated the capabilities of generative models in replicating weather phenomena at various scales, further studies are necessary to validate their reliability in capturing the physical dynamics, particularly the complex spatial variability of extreme windstorms. This study aims to address this gap by evaluating the performance of multiple generative models in simulating realistic wind fields and extreme windstorm events.

Specifically, we built four generative models, including a standard GAN, a Wasserstein GAN with gradient penalty (WGAN-GP), a U-net diffusion model, and a diffusion-GAN, to generate synthetic wind fields over the UK. These variations consist of different model architectures, training processes, and loss functions. Each model is then evaluated using multiple performance metrics to assess their ability to reproduce key characteristics of windstorm events and replicate their statistical distributions.

## 2. Data description

The dataset used in this study is the ERA5 reanalysis produced by the European Centre for Medium-Range Weather Forecasts (ECMWF). ERA5 provides hourly estimates of atmospheric variables, which covers the period from 1940 to the present with a spatial resolution of $0.25° × 0.25°$ (Hersbach et al., 2020). For this study, data from 1940 to 2022 were used to expose the models to as much training data as possible. The geographical domain focuses on the UK, covering 49°N to 59°N and 8°W to 2°E, (Fig. 1) forming a grid size of $40 × 40$. The variable used is the 10-m hourly averaged wind speed data, which captures the surface-level wind patterns and their potential impacts.

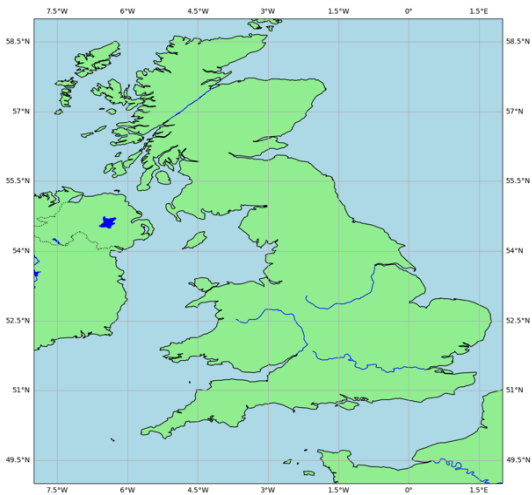

Figure 1. The spatial coverage of the ERA5 dataset used in this study and the domain for which the data samples will be generated.

Prior to model training, the ERA5 data were normalized to a range of $[0,1]$ using global minimum and maximum values across the entire domain and period. This normalization improves stability during training and generation (Glorot & Bengio, 2010). The dataset have limitations, particularly in handling surface friction due to spatial resolution constraints, which may lead to an underestimation of wind speeds over complex terrains like mountainous regions. However, the resolution remains sufficient for analyzing synoptic-scale wind patterns (Minola et al., 2020; Potisomporn et al., 2023).

## 3. Methodology

*a. Model Selection and Parameter Tuning*

Various generative models, including variational autoencoder GANs, mean and covariance feature matching GANs, and denoising diffusion GANs, were considered and tested. However, since this research serves as an exploration rather than practical application, we focused on simpler and less optimization-intensive variations that could still generate clear spatial patterns. The standard GAN and WGAN-GP were chosen as foundational models as they have shown success in simulating atmospheric conditions (Besombes et al., 2021; Brochet et al., 2023). While standard GANs can produce realistic outputs, they are prone to training instability and mode collapse. The WGAN-GP, introduced by Gulrajani et al. (2017), addresses these issues by stabilizing training through the Wasserstein distance in the loss functions, which reduces the risk of mode collapse and produces more diverse samples.

The U-net diffusion model was used for its ability to generate high-resolution outputs, while the U-net structure simplifies further tuning of the noise scheduling process. Skip connections within the U-net allows the model to retain features in various scales during training (Yao et al., 2018). Lastly, we also used the diffusion-GAN (Wang et al., 2022) as an additional choice, which combines elements of both diffusion processes and adversarial training. Although this model is relatively novel and less documented, it is conceptually a combination of the standard GAN and diffusion model. This variation is expected to combine the strengths of both approaches, which could potentially improve performance.

Given the high computational cost required to fine-tune generative models, we tuned the training parameters in a manually guided trial-and-error approach. Key parameters, such as the number of layers, learning rate, batch size, and noise scheduling, were tested and adjusted iteratively to balance training stability and performance without the need for extensive optimization. Through our experiments, we observed that the learning rate is one of the most important parameters, where a higher learning rate tended to cause overfitting and led to unphysical spatial patterns – typically characterized by large areas of unrealistically high wind over oceans, and overestimation of the storm severity index (SSI). On the other hand, a lower learning rate caused underfitting, where the models failed to capture the spatial patterns, especially in regions with complex variations.

For models that involve diffusion processes, noise scheduling was particularly important. The diffusion-GAN requires careful tuning of the noise injection process across timesteps.

Improper noise scheduling resulted in blurred patterns or failing to capture variations. In general, these generative models were found to be highly sensitive to any small changes in parameter values compared to traditional supervised learning, as there is no clear target to optimize towards. These models rely more heavily on the parameter settings, and it is more challenging to strike a balance between overfitting and underfitting.

*b. Standard Generative Adversarial Network (GAN)*

The standard GAN model comprises a generator, which produces synthetic wind fields, and a discriminator, which attempts to distinguish between the ERA5 and generated samples. The objective is for the generator to learn to produce more realistic outputs and fool the discriminator. The generator transforms a random noise vector, sampled from a normal distribution, into a $40 \times 40$ wind field through a series of transposed convolutional layers that increase the dimensionality of the data (Fig. 2). The discriminator downsamples both real and generated data through convolutional layers and learns to classify each individual sample as real or fake using a sigmoid activation function (fig. 3).

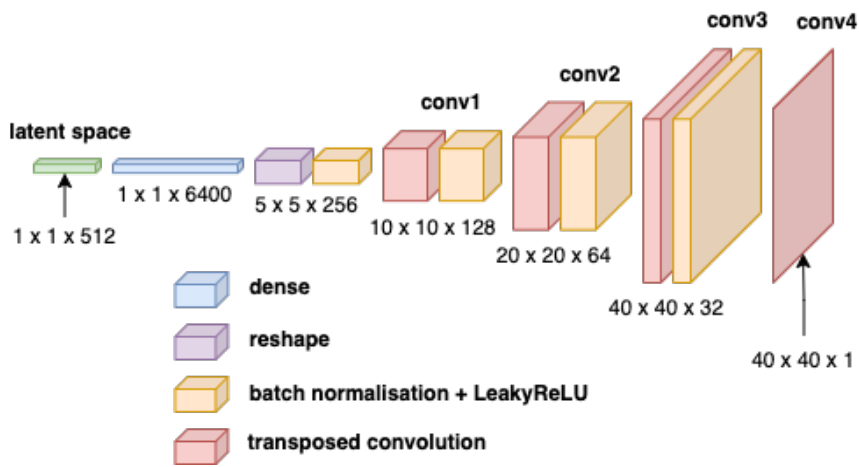

Figure 2. The architecture of the generator in the standard GAN.

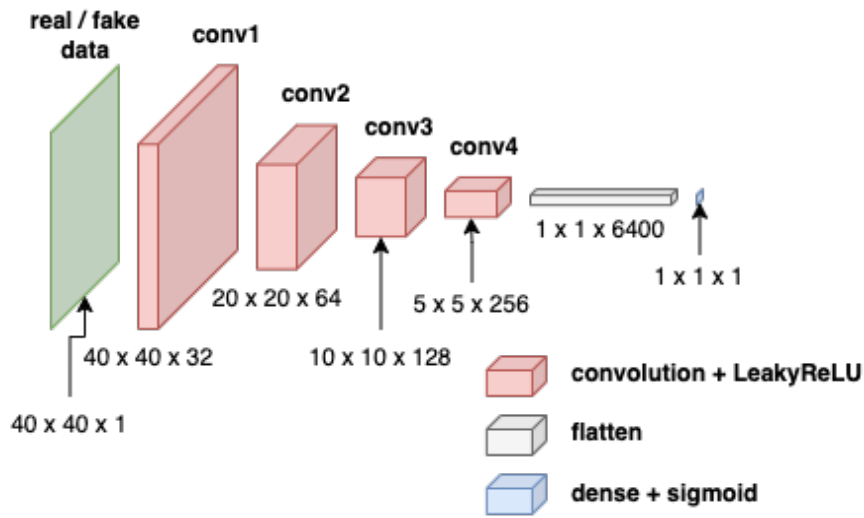

Figure 3. The architecture of the discriminator in the standard GAN.

LeakyReLU activations are used after convolutional layers in both networks, which return small values for negative inputs instead of zeros in the ReLU activation functions. This technique ensures small gradients passing through the networks and prevents vanishing gradients that may hinder the training process (Lau & Lim, 2018). Batch normalization is then applied to normalize each activation, which enhances training stability and accelerates convergence (Ioffe & Szegedy, 2015; Santurkar et al., 2018).

Both networks are trained using the adaptive moment estimation (Adam) optimizer, with learning rates of 0.0002 and batch sizes of 128 (64 real and 64 generated samples per discriminator update). These parameters are tuned iteratively, as summarized in Table 1. One-sided label smoothing is applied to the real samples, labelling them as 0.9 instead of 1 commonly found in GANs, to avoid an overconfident discriminator, and thereby vanishing gradients (Goodfellow, 2016).

| | Generator | Discriminator |
|---|---|---|
| Optimizer | Adam | |
| Learning rate | 0.0002 | |
| $\beta_1$ | 0.5 | |
| $\beta_2$ | 0.999 | |
| Loss function | Binary cross entropy | |
| Batch size | 128 | 64 real samples + 64 generated samples |
| Epochs | 10000 | |
| Real label | 1 | 0.9 |
| Fake label | | 0 |

Table 1. Optimization scheme and training parameters for the standard GAN model.

*c. Wasserstein GAN with Gradient Penalty (WGAN-GP)*

While the overall architecture of the WGAN-GP is similar to that of the standard GAN, as illustrated in Fig. 2 and 3, there are differences in the use of a critic instead of a discriminator and the addition of a gradient penalty term. Instead of labelling input sample as real or fake, the critic evaluates the realness of the samples with the Wasserstein distance. It measures the divergence between the actual and generated data distributions, which allows for a more stable training process (Shen et al., 2018). The introduction of gradient penalty ensures that the critic satisfies the Lipschitz constraint by penalizing large gradient norms that could result in poor convergence and instability. This term is computed by interpolating between actual and generated samples to ensure that the gradient norms remain close to 1, which helps provide useful gradients even when the distributions are far apart (Jolicoeur-Martineau & Mitliagkas, 2019).

Both networks are trained with the root mean square (RMSprop) optimizer, with a learning rate of 0.0006 for the generator and 0.0004 for the critic (Table 2). The critic is updated four times more frequently than the generator to provide a more accurate estimate of the Wasserstein distance before the generator updates its parameters.

|  | Generator | Critic |
|---|---|---|
| Optimizer | RMSprop | |
| Learning rate | 0.0006 | 0.0004 |
| $\rho$ | 0.9 | |
| Momentum | 0.0 | |
| Loss function | Wasserstein distance + gradient penalty | |
| Batch size | 128 | 128 real samples + 128 generated samples per update |
| Epochs | 10000 | 10000 (4 updates per epoch) |
| Real label | -1 | -1 |
| Fake label | | 1 |
| Gradient penalty weight | | 10 |

Table 2. The optimization scheme and training parameters for both the generator and critic in the WGAN-GP model.

*d. U-net Diffusion Model*

The U-net diffusion model starts with a forward diffusion process, which adds noise to the input data at each step and progressively transforms the data into pure noise. The U-net structure is then applied to learn the reverse diffusion process, by predicting the denoised versions of the input data at each step and autoregressively reconstructing the original wind fields from pure noise (Fig. 4).

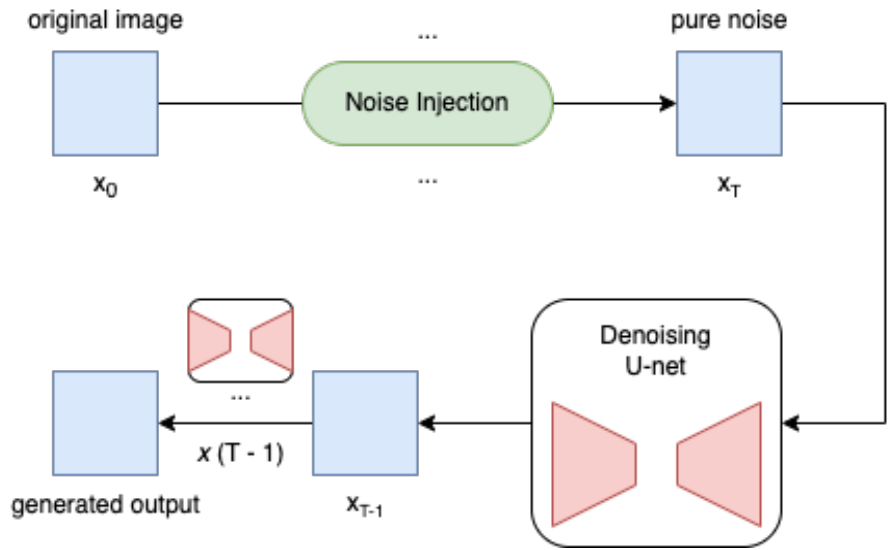

Figure 4. The overall workflow of the U-net diffusion model, including the forward diffusion (top) and the reverse diffusion (bottom) processes.

The U-net architecture (Fig. 5) consists of an encoder-decoder structure with skip connections (grey arrows) between them (Ibtehaz & Rahman, 2020). These connections ensure that the model retains important spatial features and recovers fine details in the outputs (Drozdzal et al., 2016). The encoder downsamples the noisy wind fields into lower-dimensional representations, while the decoder upsamples them into the original dimension using nearest neighbor interpolation.

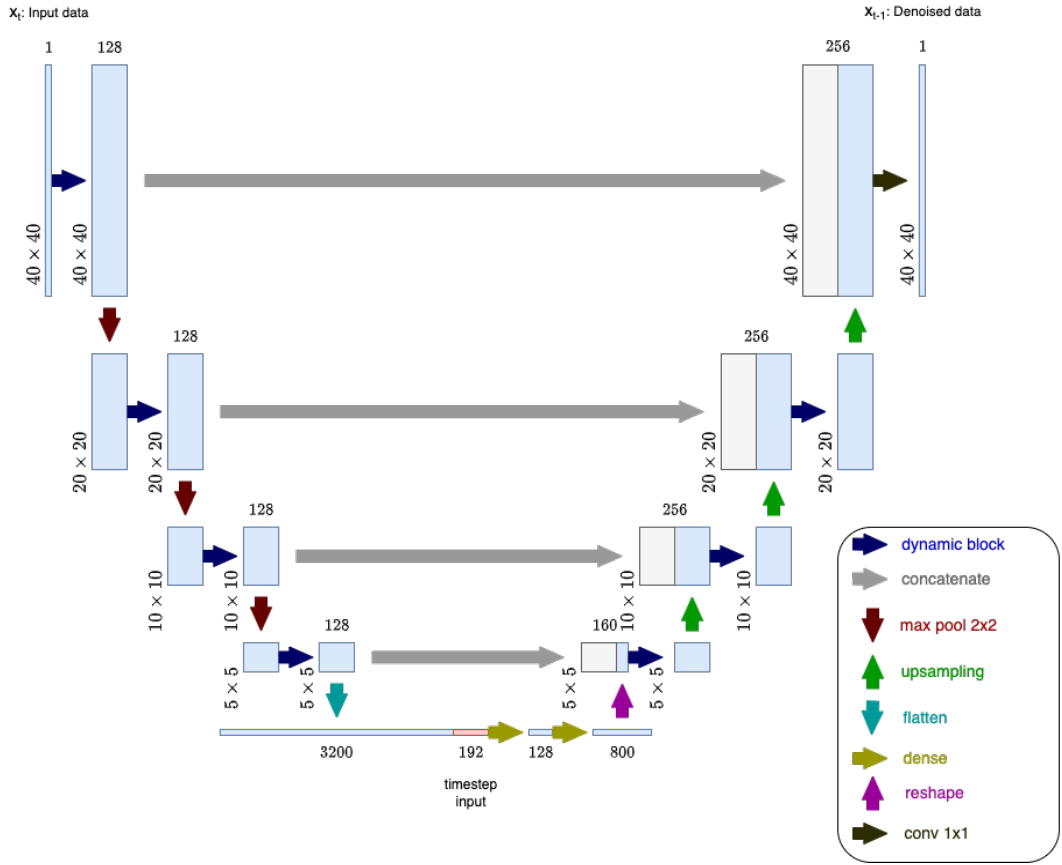

Figure 5. The architecture of the U-net used to predict the denoised data, $x_{t-1}$ given the input data $x_t$. It consists of an encoder (down-sample) and a decoder (up-sample).

In this study, the forward diffusion process follows a linear noise scheduling, where the magnitude of noise increases linearly at each timestep. If the noise is introduced too slowly, the model may fail to capture the variability in the input dataset. On the other hand, injecting noise too quickly can limit the model's ability to denoise the data effectively. The U-net structure in this study simplifies this scheduling process by allowing the mode to retain fine spatial details, even when high noise levels are injected to the wind fields.

The model is trained to minimize the mean squared error (MSE) between the denoised output and the input data at the previous timestep of the reverse process. The learning rate is progressively reduced by a factor of 0.75 if no improvement is seen after 8 epochs. This prevents the model from overfitting and allows for further tuning as the model converges (Wu et al., 2019). The model is trained for 1000 epochs, with a batch size of 128 and a linear noise schedule over 16 timesteps (Table 3).

|  | U-net Diffusion Model |
|---|---|
| Optimizer | Adam |
| Initial learning rate | 0.0008 |
| Minimum learning rate | 0.000001 |
| Learning rate reduction factor | 0.75 (after no improvement in 8 consecutive epochs) |
| $\beta_1$ | 0.9 |
| $\beta_2$ | 0.999 |
| Loss function | Mean squared error |
| Batch size | 128 |
| Epochs | 1000 |
| Batches per epoch | 569 (one-tenth of the sample size) |
| Noise schedule | Linear |
| Timesteps of noise injection | 16 |

Table 3. Optimization scheme and training parameters for the U-net diffusion model.

*e. Diffusion-GAN*

The diffusion-GAN combines the generative capabilities of traditional GANs with the noise-handling characteristics of diffusion process. Similar to GANs, this model starts with a generator that produces synthetic data from noise vectors. Gaussian noise is then progressively added to both actual wind fields from the ERA5 dataset and those created by the generator. A discriminator is designed to distinguish between these noisy versions of actual and generated data samples across multiple timesteps of the diffusion process. The inclusion of the timestep information through dynamic blocks allows the discriminator to identify realistic patterns under various noise levels (Fig. 6).

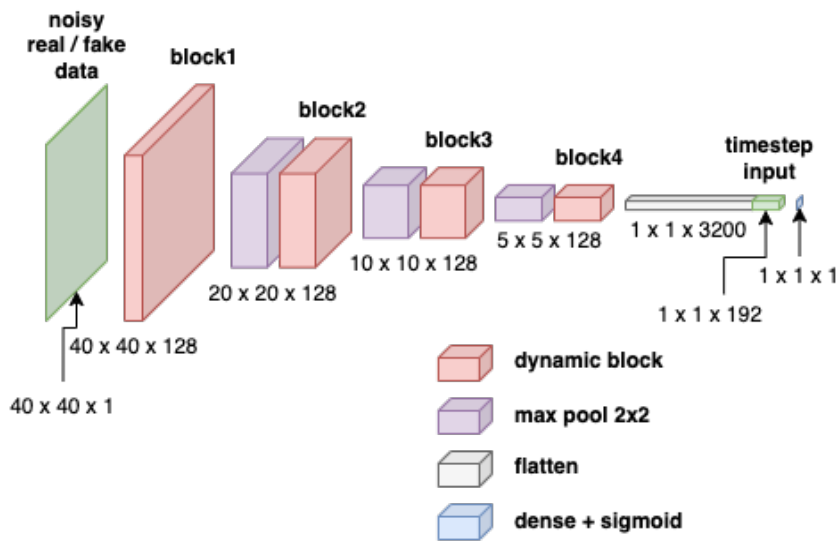

Figure 6. The architecture of the timestep-dependent discriminator in the diffusion-GAN, determines whether the inputs are the noisy versions of actual or generated samples.

Instead of a simple linear schedule used in the U-net diffusion model, the noise scheduling for the diffusion-GAN follows an exponential pattern. As the timesteps of the diffusion process increase, the proportion of the original data $\sqrt{\bar{a}_t}$ decreases progressively more rapidly, while the proportion of the Gaussian noise $\sqrt{1-\bar{a}_t}$ grows less rapidly. This schedule allows the model to better capture the variability of the data and ensures that the generator learns to produce data that remains indistinguishable at various noise levels. The noise process can be written as:

$$x_t = \sqrt{\bar{a}_t} \cdot x_0 + \sqrt{1-\bar{a}_t} \cdot \epsilon, \tag{1}$$

where $x_t$ is the noisy data at timestep $t$, $\bar{a}_t$ is the cumulative product of the noise control parameter $\alpha_t = 1 - \beta_t$, and $\epsilon$ is the Gaussian noise.

Both the generator and discriminator are trained with the same optimization scheme as in the standard GAN. The model is trained for 600 epochs, which follows the exponential noise schedule over 32 timesteps with the noise control parameter $\beta$ increasing linearly from 0.0001 to 0.02 (Table 4).

| | Generator | Timestep-dependent discriminator |
|---|---|---|
| Optimizer | Adam | |
| Learning rate | 0.0002 | |
| $\beta_1$ | 0.5 | |
| $\beta_2$ | 0.999 | |
| Loss function | Binary cross entropy | |
| Batch size | 128 | 128 real samples + 128 generated samples |
| Epochs | 600 | |
| Batches per epoch | 569 (one-tenth of the sample size) | |
| Real label | 1 | 0.9 |
| Fake label | | 0 |
| Noise schedule | | Exponential |
| Timesteps of noise injection | | 32 |
| $\beta$ in noise scheduling | | Linearly increasing from 0.0001 to 0.02 |

Table 4. Optimization scheme and training parameters for the diffusion-GAN model.

*f. Verification Methods*

The performance of the generative models was evaluated using various statistical techniques (Table 5), which focus on how well the generated outputs replicate the spatial characteristics and distributions of the ERA5 dataset. The Fréchet inception distance (FID) is a metric used to capture the similarity of generated data to the ERA5 dataset by comparing their distributions in a feature space (Heusel et al., 2017). The feature representations of the data are extracted using a pre-trained Inception-v3 model, which provides a high-level summary of the visual similarity between two datasets (Szegedy et al., 2016). A lower FID score indicates the generated wind fields resemble the actual ones in structural and statistical properties.

The FID is computed using the equation:

$$FID = \left\| \mu_r - \mu_g \right\|^2 + Tr\left( \Sigma_r + \Sigma_g - 2\sqrt{\Sigma_r \Sigma_g} \right), \qquad (2)$$

where $\mu_r$ and $\mu_g$ are the mean feature representations of the real and generated data distributions respectively, and $\Sigma_r$ and $\Sigma_g$ are the corresponding covariance matrices. The term $Tr\left( \Sigma_r + \Sigma_g - 2\sqrt{\Sigma_r \Sigma_g} \right)$ is the trace of the sum of the covariance matrices, which considers the differences in the variability and correlation structure between two distributions.

The structural similarity index measure (SSIM) was used to evaluate the spatial accuracy of the models by comparing the average storm severity index (SSI) maps. In this study, SSI quantifies the severity of windstorms by considering wind extremes, spatial extent of the storms, and the proportion of land, using the following equation (Klawa & Ulbrich, 2003):

$$SSI = \sum_{i=1}^{N} \left( LSM_i * \max\left( \frac{v_i}{v_{98,i}} - 1, 0 \right) \right)^3, \qquad (3)$$

where $LSM_i$ is the land-sea mask, $v_{98,i}$ is the local 98[th] percentile wind speed, and the sum is taken over all $N$ grid points in the domain. This ensures that only wind speeds exceeding the local extreme threshold contribute to the index.

The average SSI maps were computed across all data samples for both the ERA5 and generated datasets, identifying regions most vulnerable to wind extremes. SSIM was then used to compare these maps by assessing the magnitudes, variations, and spatial arrangements of the indices (Wang et al., 2004). A higher SSIM score indicates that the generated windstorm patterns closely replicate the spatial distribution and intensity of extreme wind events in the ERA5 dataset.

Finally, principal component analysis (PCA) was applied to reduce the dimensionality of the wind fields while preserving the most significant modes of variability (e.g., Jolliffe & Cadima, 2016). Both the Kullback-Leibler (KL) divergence and the Earth mover's distance (EMD) were computed on the distribution of the data projected onto the first 25 principal components (PCs), which account for approximately 95% of the variance in the data. The KL divergence measures the difference between the generated distribution $q(x)$ and the true one $p(x)$, while being sensitive to local discrepancies and outliers (Hu & Hong, 2013). It is computed using the equation:

$$KL(p \parallel q) = \sum_x p(x) \log\left(\frac{p(x)}{q(x)}\right).$$ (4)

On the other hand, EMD measures the minimal amount of work required to transform one distribution into another and is better for capturing global differences and overall shifts (Rubner et al., 2000). It is defined as:

$$EMD(p, q) = \inf_{\gamma \in \Gamma(p,q)} \mathbb{E}_{(x,y) \sim \gamma}[\|x - y\|],$$ (5)

where $x$ and $y$ represent individual data points in the PCA-reduced space from the ERA5 and generated datasets, respectively. The term $\Gamma(p, q)$ denotes the set of all possible joint probability distributions $\gamma(x, y)$ that have $p$ and $q$ as their marginal distributions, while the expected value $\mathbb{E}_{(x,y) \sim \gamma}[\|x - y\|]$ represents the average distance between pairs of points $x$ and $y$, averaging over the joint distribution $\gamma$. The infimum (inf) represents the minimal value of the expected distance overall over possible joint distributions $\gamma$.

| Metrics | Description | Importance |
|---|---|---|
| Fréchet inception distance (FID) | Measure the similarity between distributions in a feature space | Evaluates overall quality of generated outputs |
| Structural similarity index measure (SSIM) | Measures the spatial similarity of extreme wind events | Assesses structural accuracy in extreme events |
| Kullback-Leibler divergence (KL divergence) | Measures the divergence in the first 25 PCs | Captures fine-scale discrepancies |
| Earth mover's distance (EMD) | Measures the distributional similarity in the first 25 PCs | Captures the global distributional differences |

Table 5. Summary of the evaluation metrics.

# 4. Results

## a. Visual Comparison

Fig. 7 shows a random selection of typical 10-m wind fields from ERA5 and the four generative models. As it depicts 'typical' days, the wind speeds shown are generally much lower than those observed during windstorm events. We observe that ERA5 (Fig. 7a) generally shows higher wind speeds over open waters such as the North Sea and the English

Channel and slightly higher wind speeds in coastal regions. Blue colours, indicating lower wind speeds, typically below 10 m s$^{-1}$, are more common inland.

All four generative models (Fig. 7b-e) show similar spatial patterns of wind speed across the UK, indicating that they can produce reasonably realistic winds for typical days. However, there are some slight differences observed between different models, particularly in image quality. The standard GAN, while replicating the overall spatial structure, tends to produce slightly blurry wind maps with less smooth gradients and boundaries, particularly over oceans. The U-net diffusion model, on the other hand, generates the least noisy outputs similar to the ERA5. The other outputs appear realistic in general, with no significant differences observed.

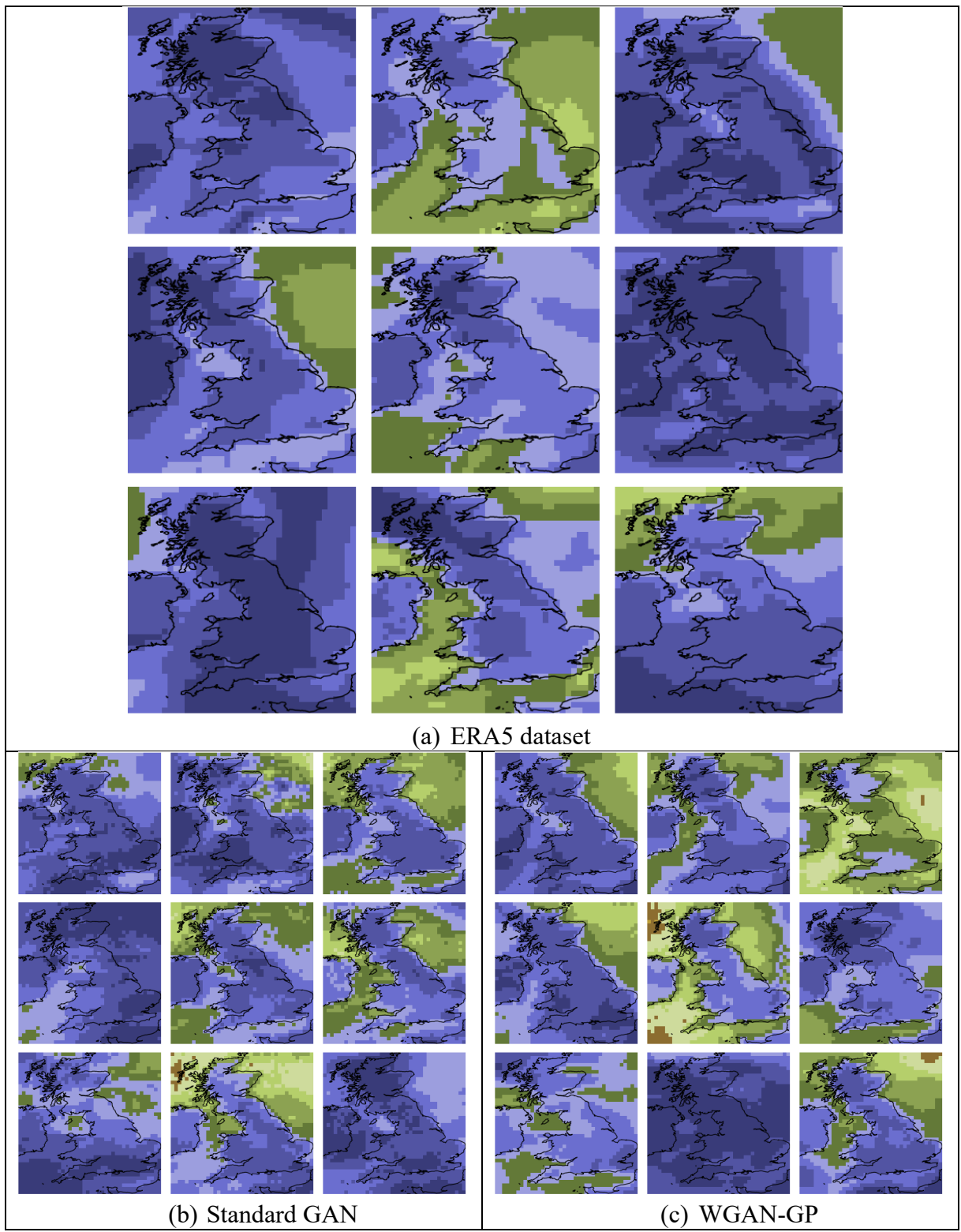

(a) ERA5 dataset

(b) Standard GAN          (c) WGAN-GP

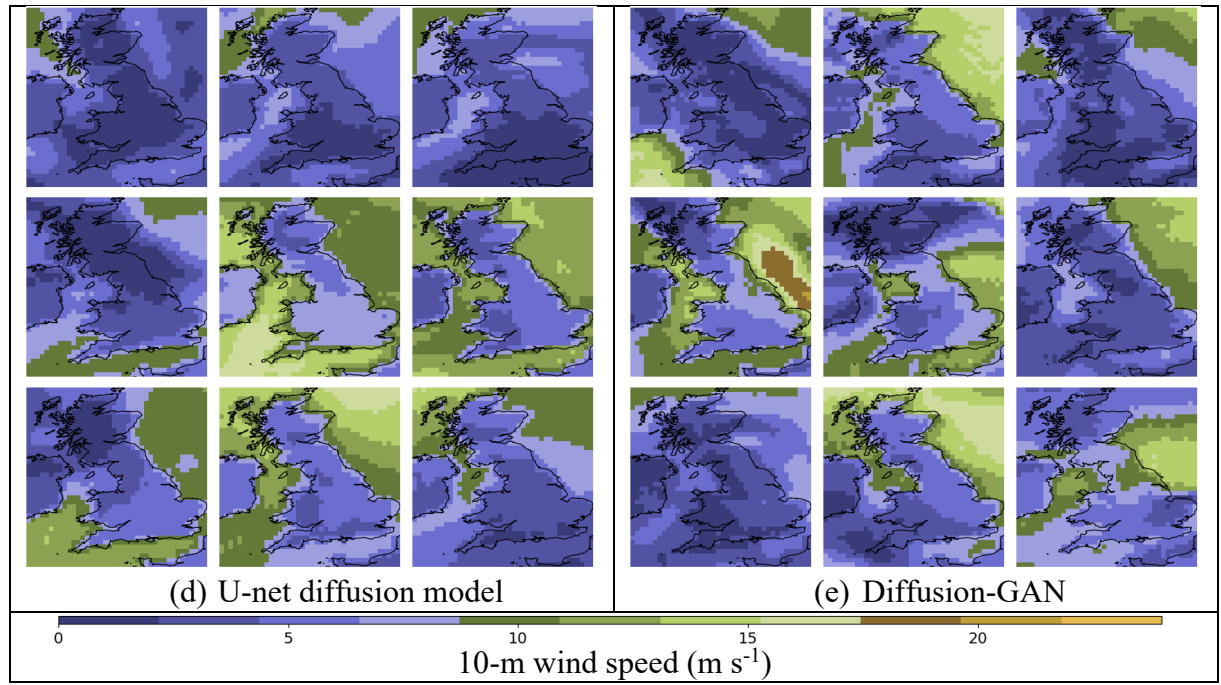

Figure 7. Comparison of typical wind speed maps. Each box shows nine randomly chosen 10-metre hourly-averaged wind speed maps (m s⁻¹) from the ERA5 dataset (a), standard GAN (b), WGAN-GP (c), U-net diffusion model (d), and diffusion-GAN (e).

To investigate whether the generative models can reliably produce extreme cases, we ran the models to produce the same number of samples as the ERA5 and compared the top 10 SSI events in these datasets (Fig. 8), with individual storm intensity reflected by the SSI value above each map. We observed that the ERA5 dataset (Fig. 8a) shows wind speeds exceeding 20 m s⁻¹ in certain regions, particularly over the Irish Sea and English Channel, and occasionally over the North Sea and near coastal regions. These extreme wind events show well-defined footprints with clear differences between areas of high and low wind speeds, with wind speeds below 10 m s⁻¹ commonly seen inland in Scotland and central England.

The standard GAN (Fig. 8b) sometimes overextends the regions of moderate and high wind speeds, covered by red colors on the maps, particularly over the North Sea and Scottish coast. This results in less clear footprints and slight overestimation of the SSI values of such extreme cases, where they are slightly larger than the corresponding values from the ERA5 dataset. This may be due to overfitting, leading the model to overemphasize the differences between inland and coastal regions. As a result, the spatial distribution of wind speeds produced by the standard GAN does not match the input dataset as closely as the other models.

The WGAN-GP (Fig. 8c) model also tends to overextend the regions of high wind speed but produces a similar range of SSI values among the top 10 SSI cases. There are subtle differences in where high wind speeds are more frequently observed. Some output maps place the areas of strongest winds over the North Sea or Scottish coast, rather than the Irish Sea or the English Channel commonly seen in the ERA5 dataset. Additionally, some outputs from the WGAN-GP show some localized random impulse noise near the eastern boundary of the domain, which affects the overall image quality and smoothness of the maps.

The U-net diffusion model (Fig. 8d) produces very realistic structures, with smooth transitions between wind speeds that closely resemble the ERA5 dataset. However, it significantly underestimates the SSI values of extreme wind events. The third highest SSI value from this model is 50.98, which is significantly lower than any top 10 cases in the ERA5 dataset. The model tendency to smooth out high wind speeds helps produce clearer footprints and visually coherent outputs but might make it less effective in capturing the intensity of extreme windstorms.

Finally, the diffusion-GAN (Fig. 8e) also creates similar spatial patterns to the ERA5 dataset, with occasional lower wind speeds found over Scotland and other inland areas. The overall image quality is slightly lower than the U-net diffusion model, with less smooth boundaries between the high and low winds. The model also overestimates the storm intensity, with the top SSI value reaching 181 and the top 6 SSI values exceeding 100. Despite these discrepancies, the U-net diffusion model and diffusion-GAN appear to have the best performance in replicating the spatial characteristics of extreme wind events.

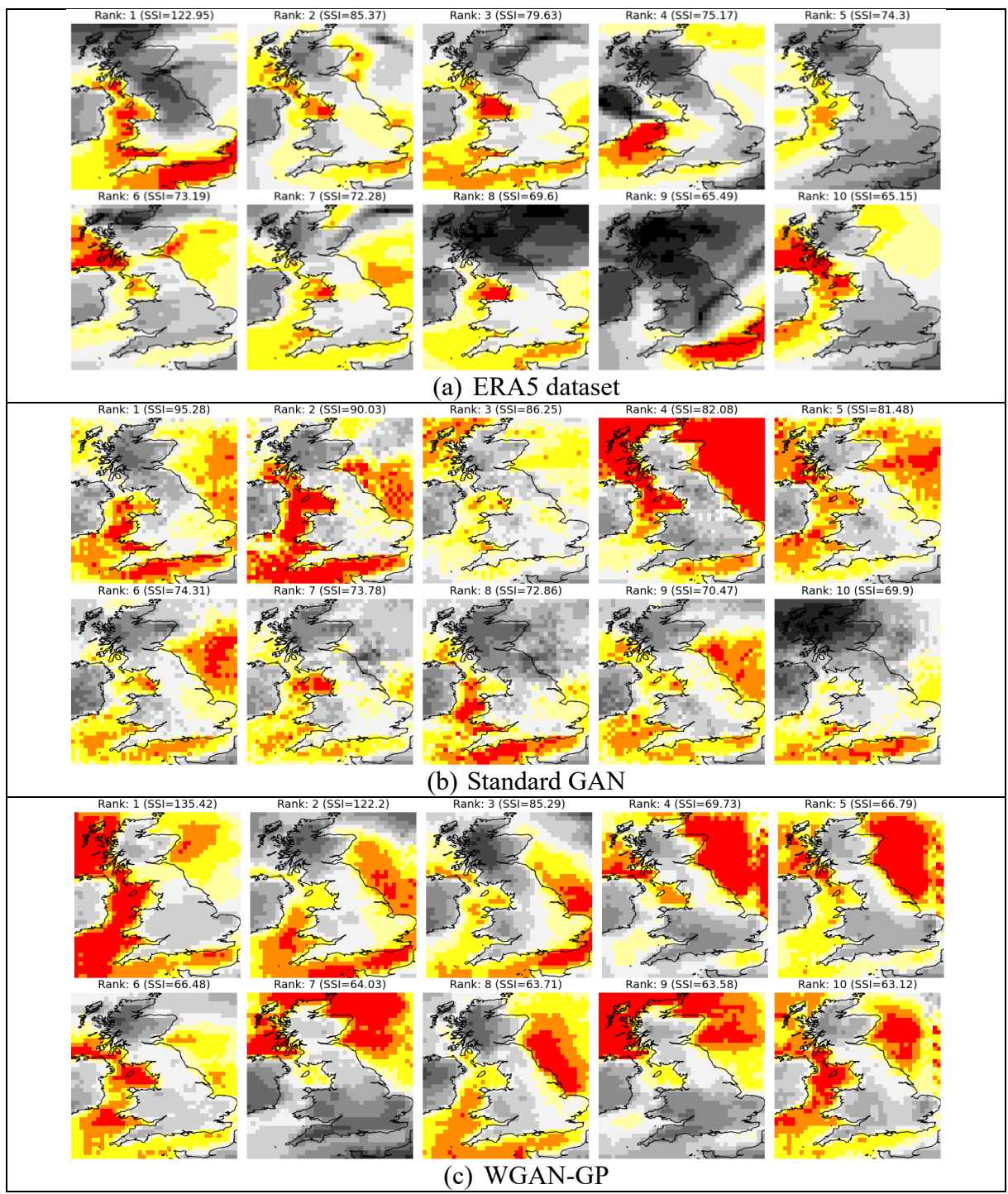

(a) ERA5 dataset

(b) Standard GAN

(c) WGAN-GP

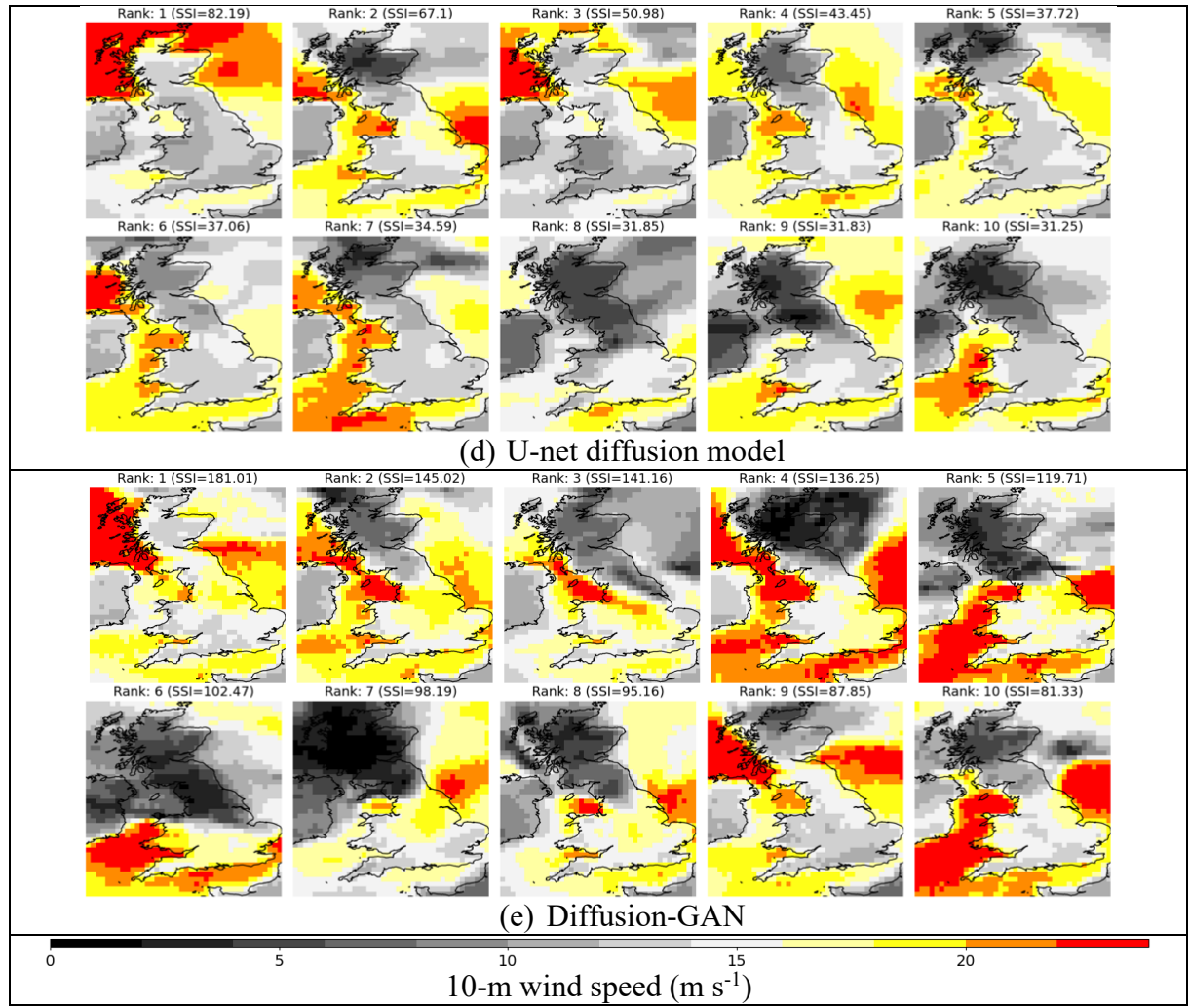

Figure 8. Comparison of extreme wind speed maps. Each box shows the 10-metre wind speed maps in metres per second (m s$^{-1}$) of the top 10 SSI cases, from the ERA5 dataset (a), standard GAN (b), WGAN-GP (c), U-net diffusion model (d), and diffusion-GAN (e).

### b. SSI Distribution Analysis

The SSI distribution (Fig. 9) shows a steep decline in frequency as SSI values increase across all models and the ERA5 dataset (blue line). The U-net diffusion model (red line) substantially underestimates the intensity of such wind events, failing to match the frequencies of high-intensity wind events in ERA5. The distributions of the other models align more closely with the ERA5 dataset, starting from a steep drop to a gradual decline. Among them, the WGAN-GP (green line) and diffusion-GAN (purple line) show closer alignment with the ERA5 dataset at lower SSI values from 10 to 30, while all models underestimate the frequency of extreme events (SSI values above 60). This underrepresentation in the tail of the distribution may reflect the inherent tendency of generative models to generalize patterns from the training data. This behaviour often results in the generation of more samples within well-documented ranges, while rare extremes are less emphasized during training.

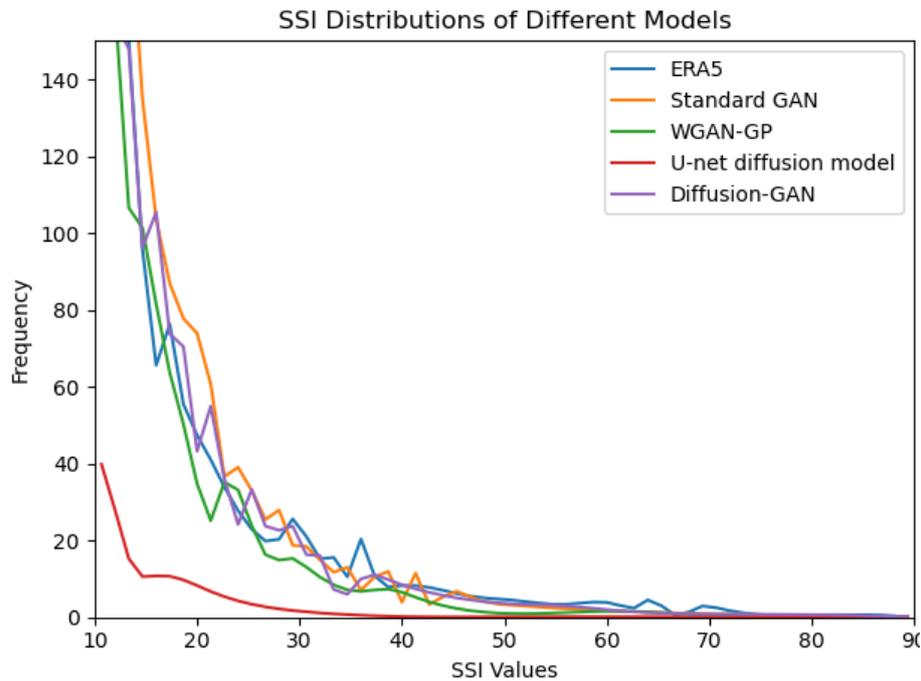

Figure 9. SSI distribution plot showing the frequency of SSI values of individual wind maps for extreme windstorms, comparing the ERA5 dataset (blue line) with the outputs from the standard GAN (orange line), WGAN-GP (green line), U-net diffusion model (red line), and diffusion-GAN (purple line).

*c. Principal Component Analysis (PCA)*

The empirical orthogonal functions of the first two principal components (PCs), contributing around two-thirds of the variability in ERA5, help describe the spatial characteristics of observed windstorm patterns. The PC1 loading (Fig. 10a) reflects a mean intensity component, reflected by the negative skewness in the PC1 distribution and larger PC1 values among the top 100 SSI cases (Fig. 11a). On the other hand, the PC2 loading (Fig. 10b) shows a north-south gradient pattern. The variability shown in PC2 values among the top 100 SSI cases (Fig. 11a) suggests that PC2 likely captures a geographical component, representing the latitudes at which these storms would impact the UK.

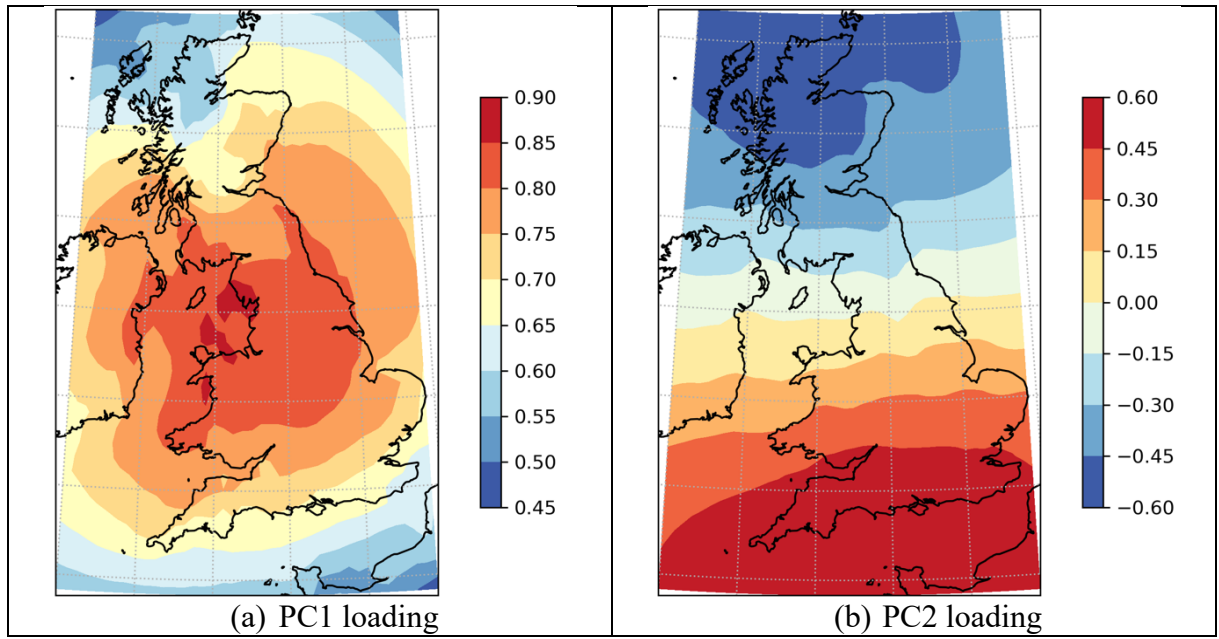

Figure 10. Standardized spatial loadings of the first two PCs for 10-m wind fields in the UK. Units are arbitrary.

The filled contours in Fig. 11 represent the logarithmic frequency of occurrences, with darker regions indicating higher frequency. The top 100 SSI cases in each model are shown in blue dots, characterising how well each model replicates the distribution of extreme wind events in the reduced dimensional space. The red dotted lines represent the PC bases, which indicate the directions in which the variance in the data is maximized. In ERA5 (Fig. 11a), the contours are concentrated near the center, which extend along the direction of the PC1 vector and display a negative skew. The top 100 SSI cases are mostly distributed around high values of PC1, with slightly more located in the positive (i.e. over southern England) than the negative side (i.e. Scotland) of the PC2 dimension.

The four models (Fig. 11b-e) show similar shape and density of contours, but with slight variations, particularly in the distribution of the extreme cases. All four models show a more dispersed distribution of the top 100 SSI cases on the PC dimensions, with the blue scatter points located further away from the ERA5 cluster. This increased dispersion suggests that the models introduce more variability in the extreme wind events than the ERA5 dataset. Specifically, for both the standard GAN and WGAN-GP, the extreme cases tend to have slightly more negative PC2 values. This suggests a potential bias in these models towards more storms near the Scottish coast than reality. Additionally, the PC1 vector in the standard GAN shows a slight anticlockwise rotation compared to the ERA5 dataset and the other models. This indicates a shift in how this model generates the primary variance in the dataset and might be indicative of instability.

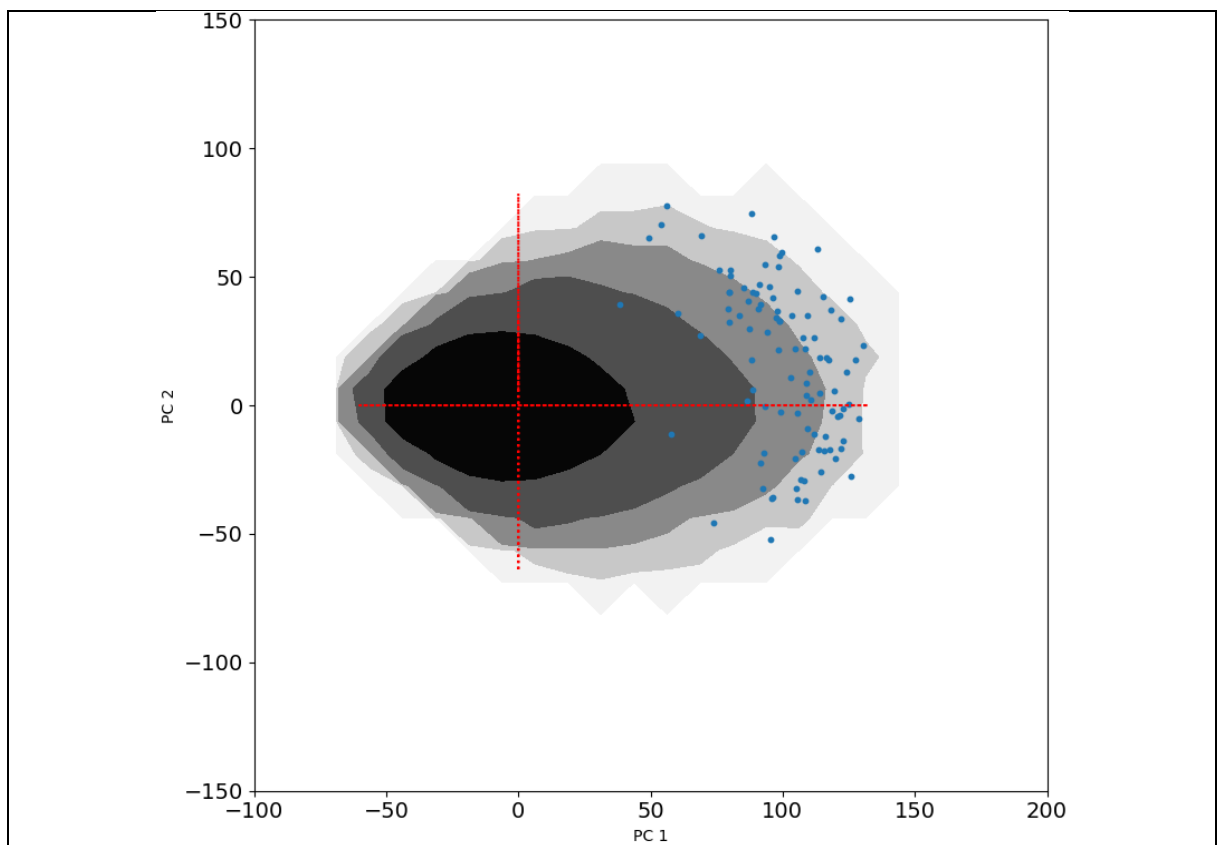

(a) ERA5 dataset

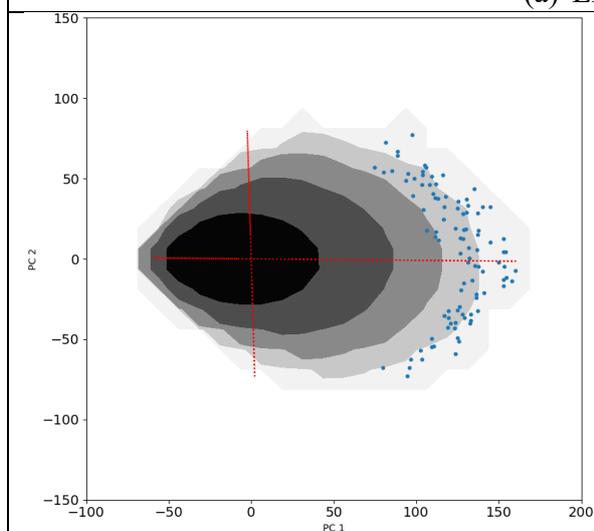

(b) Standard GAN

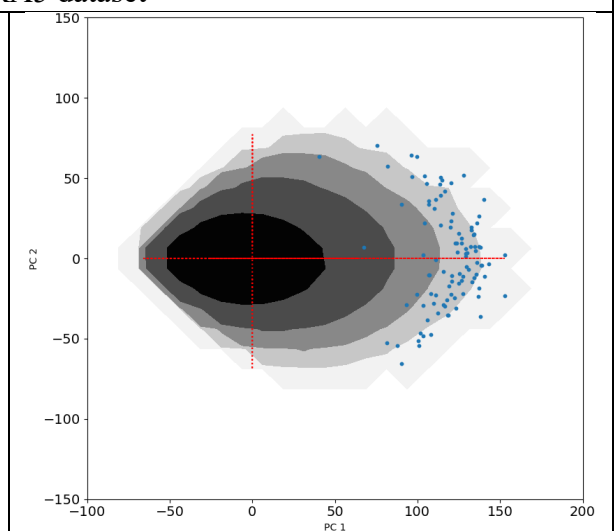

(c) WGAN-GP

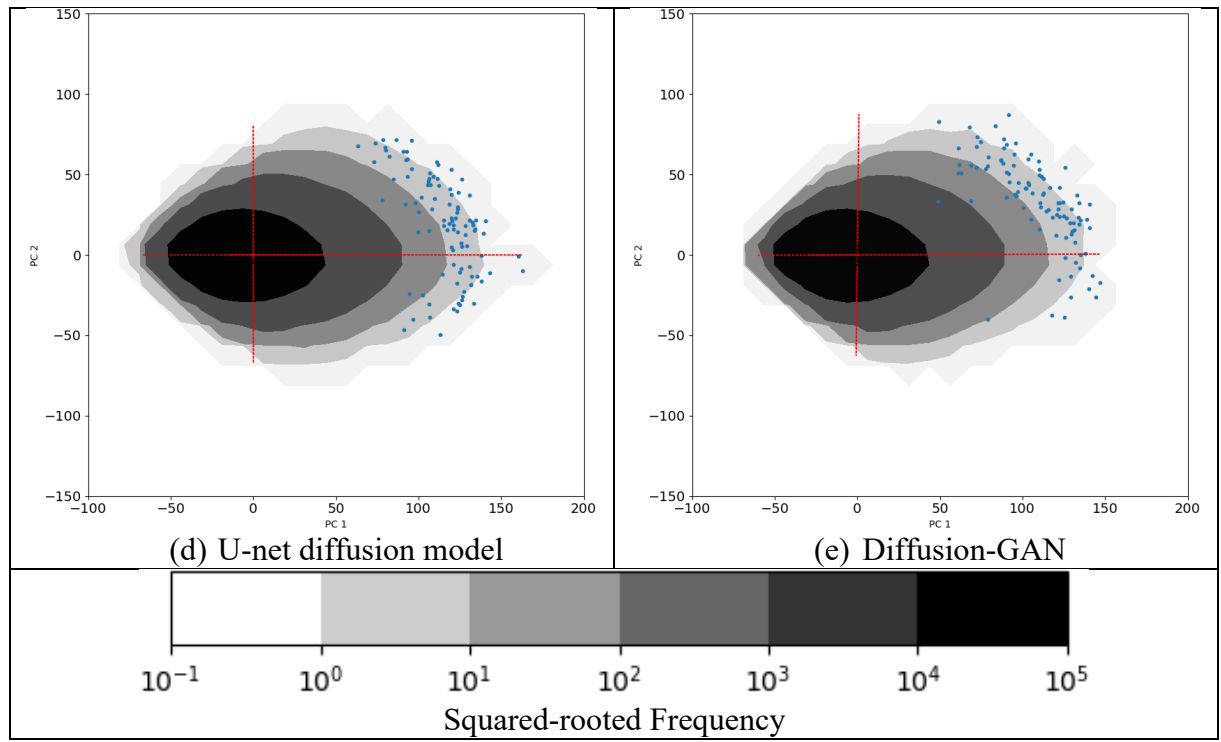

Figure 11. The PCA plots show the distribution of the ERA5 dataset (row 1), standard GAN (row 2, left), WGAN-GP (row 2, right), U-net diffusion model (row 3, left), and diffusion-GAN (row 3, right) along the first two PC dimensions. The contour lines represent the square-rooted frequency of occurrences. The red dotted lines represent the PC vectors (eigenvectors). The blue scatter points represent the top 100 SSI cases.

*d. Point-based Analysis*

London, Edinburgh, the Irish Sea, and the North Sea are selected to represent different geographical settings with varying exposures to strong winds. London and Edinburgh are urban areas, while the Irish Sea and the North Sea are open water regions, each with different exposures to strong winds and extreme windstorms. The boxplots of wind speed distribution at these four locations (Fig. 12) help assess how each model captures the wind distributions across different environments and provide insights into their strengths and weaknesses.

In London (Fig. 12a), all models generally capture the bulk of the wind speed distribution when compared to the ERA5 dataset. The WGAN-GP produces a slightly higher median wind speed compared to the other models. The interquartile range (IQR) remains similar across most models, with the U-net diffusion model showing a narrower IQR and a smaller upper quartile, indicating its tendency to underestimate higher wind speeds. The whiskers, representing the 1st and 99th percentiles, show that all models capture the typical wind speeds. However, the lower wind speeds in the outliers above the upper whisker across all models compared to the ERA5 dataset, suggest that they struggle in producing the most extreme values over London. This aligns with the findings from the SSI distribution analysis (Fig. 9), where all models underrepresent extreme values.

Over Edinburgh (Fig. 12b), the median wind speeds are similar across all models, but their distributions show more variation. The WGAN-GP produces a wider range of typical wind speeds, indicated by the wider IQR and whiskers. The diffusion-GAN captures the overall distribution well, with very similar extreme values found above the upper whisker. On the other hand, the other three models show lower values in the outliers, suggesting that they again represent less extreme wind speeds over Edinburgh compared to ERA5.

Over the Irish Sea (Fig. 12c), wind speeds are generally higher than over land, with relatively smaller extreme values compared to the 1$^{st}$ percentile. The IQRs of the WGAN-GP and diffusion-GAN show a similar spread compared to ERA5, suggesting that these models generate a realistic range of typical wind speeds. However, the standard GAN and U-net diffusion model have narrower IQRs than ERA5, indicating that they produce wind speeds more clustered around the median at this location. Like Edinburgh, fewer extreme outliers are found above the upper whisker for the standard GAN and U-net diffusion model, while the WGAN-GP and diffusion-GAN represent the extreme values found in ERA5.

In the North Sea, the standard GAN model shows slightly larger extreme values above the upper whisker, indicating that it generates higher wind speeds more frequently than observed. Similarly, the U-net diffusion model shows a narrower IQR and fewer extreme outliers, which suggests that the model produces a more constrained range of wind speeds. The less extreme outliers in this model also indicate a tendency to underestimate variability at the upper bound while producing fewer extreme outliers when compared to the ERA5 dataset and the other models.

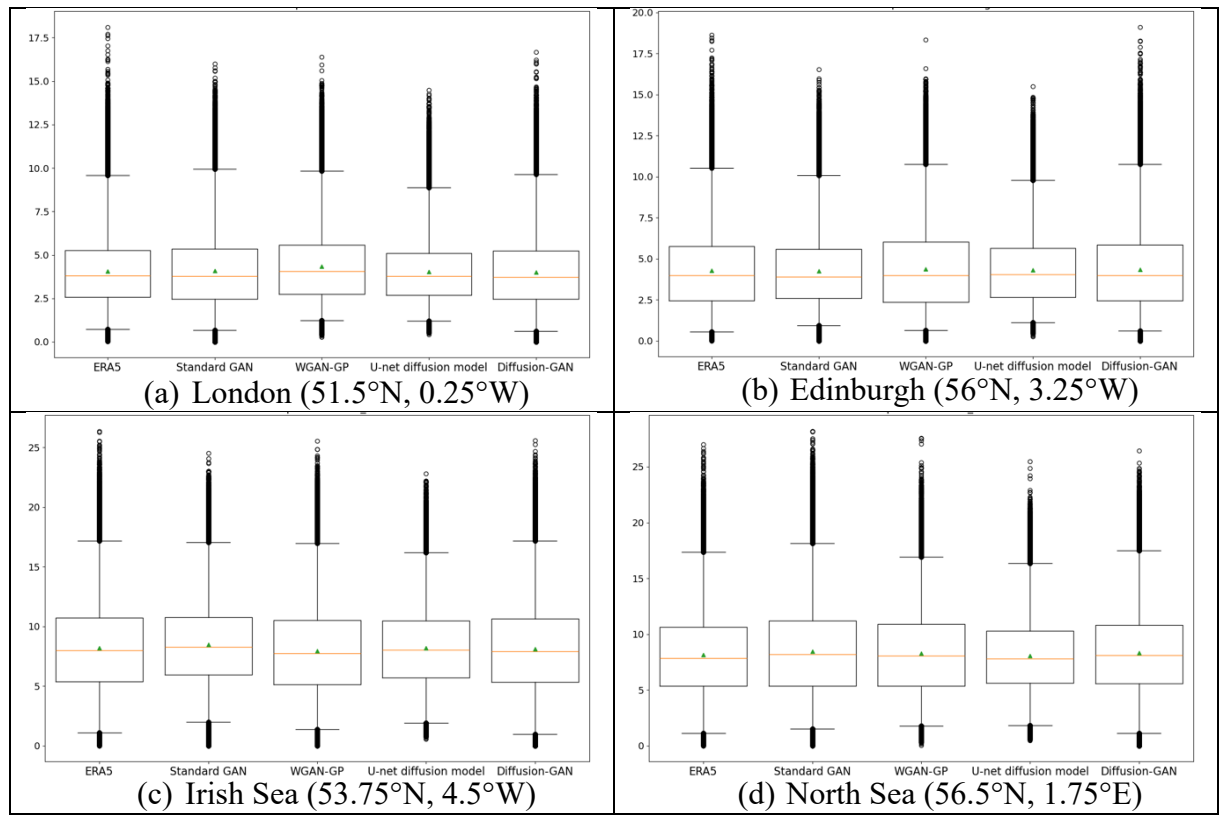

Figure 12. Boxplots of 10-m wind speed distributions at four selected locations, London (a), Edinburgh (b), the Irish Sea (c), and the North Sea (d). They compare the ERA5 dataset (column 1) with the outputs from the standard GAN (column 2), WGAN-GP (column 3), U-net diffusion model (column 4), and diffusion-GAN (column 5). The boxes represent the interquartile range (IQR) from the lower quartile (Q1) to the upper quartile (Q3), with the median wind speed indicated by the horizontal orange line. The whiskers extend to the 1st and 99th percentiles, capturing the typical wind speed range. Outliers above the whiskers are shown as individual dots that represent extreme wind speeds.

By analyzing the return period plots at the same four locations (Fig. 13), we focus on the tail-end distributions of wind speeds captured by the models. Across all locations, the U-net diffusion model (red line) consistently overestimates the return periods of moderate to high wind speeds (10-30 m s$^{-1}$). This indicates the model's tendency to underestimate the frequency of such events, as suggested by its underestimation of wind speeds from the previous analysis. For the other models , at London, Edinburgh, and the Irish Sea (Fig. 13 a-c), the distributions align closely with the ERA5 for much of the range but commonly show slight overestimations of return periods at the rarer tail-end (extreme wind events with return periods above 10$^3$ days).

In the North Sea (Fig. 13d), the standard GAN (orange line) underestimates the return periods across the displayed wind speed range. On the other hand, the WGAN-GP (green line) and diffusion-GAN (purple line) align well with the ERA5 with slight overestimations at the rarer extremes (return periods above 10$^3$ days). This consistency in capturing the distributions reflects the ability of the two models to accurately simulate distributions of wind containing moderate to high wind speeds, with the spread in the tail expected due to the inherent variability in modelling and observing such rare events.

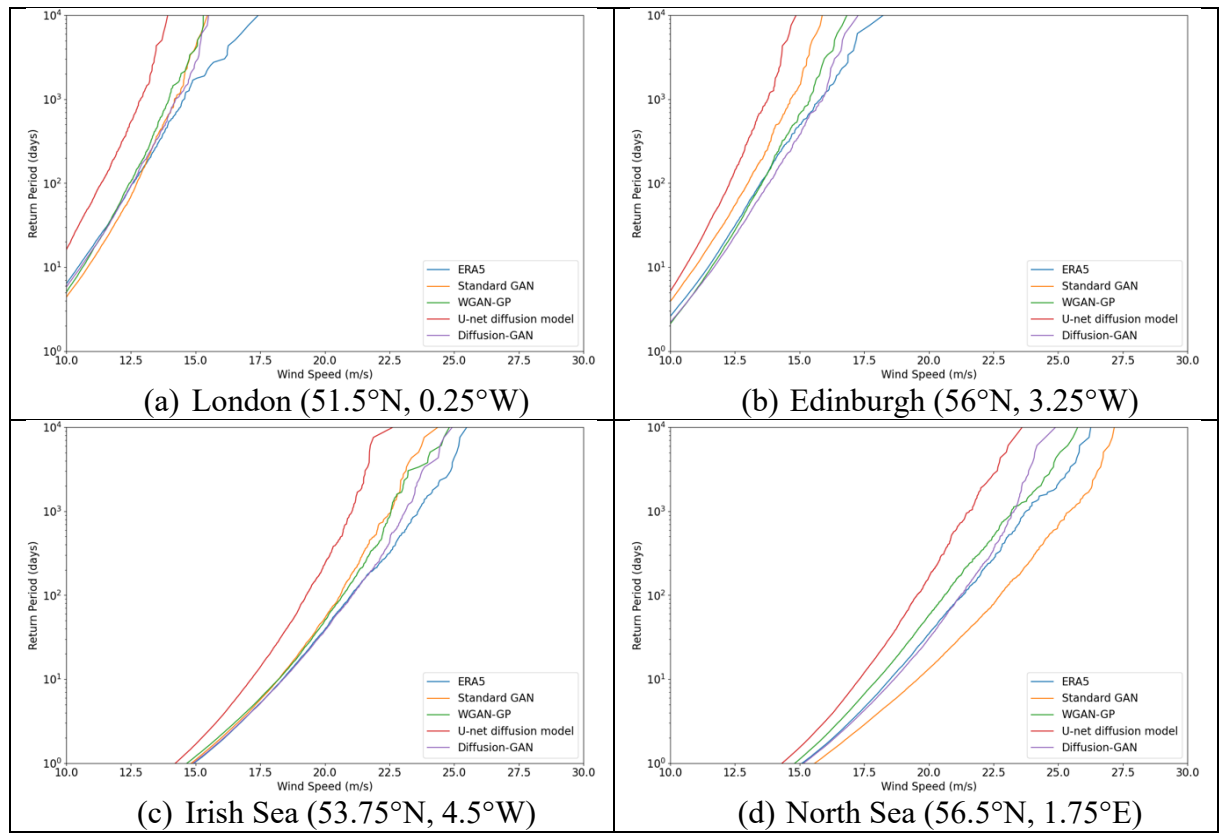

Figure 13. Return period (in days) of wind speeds ranging from 10 to 30 m s⁻¹ at London (a), Edinburgh (b), Irish Sea (c), and North Sea (d), comparing the ERA5 (blue) with the standard GAN (orange), WGAN-GP (green), U-net diffusion model (red).

*e. Performance Evaluation*

The model performance is further evaluated using statistical metrics (Table 6). We observed that the gaps between models for FID and SSIM are relatively small, while the differences in KL divergence and EMD are larger. This suggests that there is only a minor gap in how these models align with ERA5 in feature space and capture the spatial distributions of extreme wind events. On the other hand, the larger gaps in KL divergence and EMD indicate greater differences in the model ability to match the probability distribution and variability of the wind patterns in the PCA space.

The standard GAN ranks second in FID and first in SSIM, suggesting it captures the overall patterns of extreme events slightly more effectively, with strong alignment in visual quality to the ERA5 dataset. However, in terms of KL divergence, the standard GAN has the highest value of 0.910, which is significantly larger than those of the other models. It also ranks third in EMD with 0.207 and is more than three times higher than the top-performing model. These results indicate that the standard GAN diverges more from the ERA5 dataset in terms of the distributional properties in the reduced-dimensional space.

The WGAN-GP ranks more evenly across all metrics, being second in SSIM, KL divergence, and EMD. This balanced performance suggests that the model performs relatively well in capturing multiple statistical properties. The second-place ranking in KL divergence and EMD indicates a strong performance in replicating both fine-scale and global distributional characteristics of the distributions present in the ERA5.

The U-net diffusion model ranks first in FID, showing that it captures the overall visual quality better than the other models. However, it ranks third in KL divergence, and the last in SSIM and EMD. These results suggest that the U-net diffusion model underperforms in capturing the structural details of extreme events and matching the distribution of wind patterns in reduced-dimensional space. These lower rankings could be potentially due to a significant underestimation of extreme wind intensities, limiting its accuracy in capturing the statistical distributions and representing regions prone to intense wind events.

The diffusion-GAN ranks fourth in FID and third in SSIM, indicating that it performs slightly less well in capturing the visual quality and structural similarity of extreme events compared to the other models. However, the model performs significantly better in KL divergence and EMD, suggesting that it closely replicates the distributional characteristics of the ERA5 dataset in the reduced dimensional space, in both fine-scale and global distribution matching.

| Metrics / Datasets | FID | SSIM | KL Divergence | EMD |
|---|---|---|---|---|
| ERA5 | 0.000 | 1.000 | 0.000 | 0.000 |
| Standard GAN | 721 (2) | 0.162 (1) | 0.910 (4) | 0.207 (3) |
| WGAN-GP | 728 (3) | 0.155 (2) | 0.525 (2) | 0.138 (2) |
| U-net diffusion model | 717 (1) | 0.151 (4) | 0.613 (3) | 0.225 (4) |
| Diffusion-GAN | 729 (4) | 0.154 (3) | 0.488 (1) | 0.067 (1) |

Table 6. Evaluation metrics for each model, including Fréchet inception distance (FID), structural similarity index measure (SSIM), Kullback-Leibler divergence (KL divergence), and Earth mover's distance (EMD). The numbers in brackets represent the rank of each model for each metric, with (1) indicating the best performance and (4) the worst.

# 5. Discussion

## a. Key Findings

While all models successfully captured the general wind behaviors across different regions, there are differences in terms of statistical agreement and representation of extremes. As a baseline model, the standard GAN performed adequately overall while exhibiting limitations in replicating the intensity and variability of extreme events. The WGAN-GP captured the intensity slightly better than the standard GAN but still occasionally overextended regions of extreme wind speeds. Both models also struggled to maintain image quality, usually producing blurry outputs with noise on top of the accurate structures.

In contrast, the U-net diffusion model consistently produced the highest visual quality with the least noise. However, it often underestimated extreme wind speeds, as shown from the SSI distributions and point-wise wind speed comparisons. The model architecture appears to over-generalize the spatial patterns during learning, resulting in more stable but less extreme outputs, i.e. underestimating the population variance. On the other hand, the diffusion-GAN generated higher wind speeds and SSI values that aligned closely with extreme events in the ERA5 dataset. However, the most extreme values of SSI in the distribution are much higher than those seen in ERA5.

These findings suggest that model architectures and training strategies played a key role in model behavior. For example, the diffusion process in the U-net diffusion model and diffusion-GAN appeared to stabilize outputs and improve image quality. On the other hand, the gradient penalties in the WGAN-GP seemed to contribute to better statistical agreement but did not fully mitigate the weaknesses of the standard GAN, such as the overextension in high-intensity regions and noisier outputs.

## b. Variability across Outputs

Another key aspect explored in this study is the variability in outputs generated by the models across multiple runs. The spread of SSI distributions (Fig. 14) shows how different models respond to 20 repeated individual samplings under the same settings. The U-net diffusion model (Fig. 14c) consistently underestimates SSI values, with the ERA5 distribution (blue line) lying above its maximum outputs across all ranges.

The other three models (Fig. 14 a-b, d) also produce relatively consistent outputs, with the ERA5 distribution generally within the range of generated outputs. Among them, the standard GAN (Fig. 14a) shows the broadest range of variability, with the ERA5 distribution being closer to its minimum for lower SSI values and moving to the median in the upper range. The WGAN-GP (Fig. 14b) consistently places the ERA5 distribution near its maximum across all ranges, though it still underestimates the frequency of extremes with SSI greater than 60. On the other hand, the diffusion-GAN (Fig. 14d) shows the best performance, with the ERA5 distribution typically near the median. These findings suggest an ensemble approach to address individual limitations and potentially improve overall generative capabilities across different windstorm scenarios.

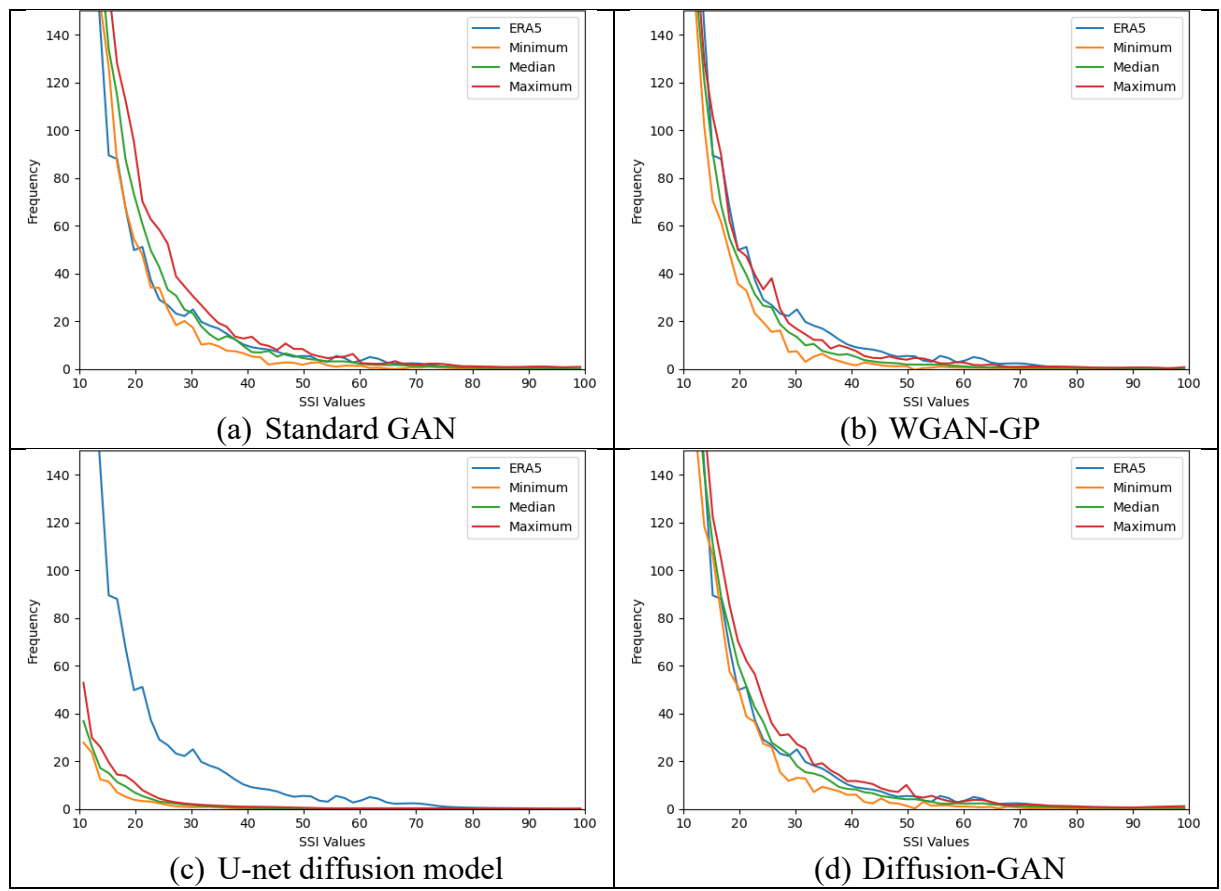

Figure 14. Spread of SSI distributions for each model across 20 generated datasets of equal length (83 years), comparing the ERA5 dataset (blue line in each plot) to the standard GAN (a), WGAN-GP (b), U-net diffusion model (c), and diffusion-GAN (d) respectively. The range of SSI values generated by each model is represented with the minimum (yellow), median (green), and maximum (red) frequency across all datasets.

*c. Limitations*

There are inherent limitations in the ERA5 dataset due to its spatial resolution and potential biases, particularly over complex terrains like mountains in Scottish Highlands. These constraints may lead to inaccuracies in capturing localized windstorm characteristics. Additionally, the rarity of tail-end extreme events in the training dataset limits the models' training exposure to the full spectrum of windstorm intensities, particularly high-impact scenarios. The use of historical reanalysis data also limits the applicability of such models for assessing future windstorm patterns, as they are unable to learn shifts in windstorm behavior under climate change.

On the other hand, the generative models in this study are limited in capturing the full physical dynamic of windstorms. As these models rely solely on statistical patterns within the training dataset, the outputs may sometimes lack physical realism, such as the U-net diffusion model's tendency to produce overly generalized outputs and WGAN-GP's overextension of intense wind regions. Moreover, the models generate independent and identically distributed (IID), single-variable fields without temporal information or relationships with other meteorological variables. While this design simplifies the training process, it restricts the models' ability to represent the dynamic evolution of windstorms. Future work could explore the use of temporal information and other meteorological fields in producing more physically robust windstorm populations.

Additionally, the selection and optimization of models in this study relied on a trial-and-error approach instead of hyperparameter tuning due to computational complexity. As a result, model performance is not fully optimized and left room for improvement. Finally, the evaluation of models in this study relies on metrics such as FID and SSIM, which do not fully capture the spatial complexity and physical characteristics of windstorms. These metrics primarily assess statistical alignment but do not notice subtler, meteorologically important structural differences.

*d. Future Directions*

Incorporating multiple meteorological variables and temporal dimensions could enhance the model performance and their applicability. For example, outputs from RCMs would allow the models to simulate future windstorm patterns under climate change, which could improve future risk assessment. Generating extensive synthetic datasets from various historical periods could also help understand the role of climate change in windstorm patterns and enable more robust analyses in windstorm impact. Temporal information could be integrated through generating windstorm footprints that capture the maximum wind speeds over a given period or producing multi-frame outputs that represent wind field evolution over time. Additionally, the domain size in this study is relatively small. Future research could explore the potential of applying similar methods to larger domains like Europe. This approach could also be expanded to other meteorological hazards, including tropical cyclones and severe convective systems, as well as other regions, such as the United States and Japan. These extensions could broaden the applicability of generative models to diverse hazards and regional climates.

Further studies could also focus on understanding the strengths of different models and their applicability to different scenarios. For instance, analysis could identify which models perform better in specific regions or for certain storm characteristics and intensities. In this case, an ensemble approach could then combine these strengths for targeted applications, with optimization performed based on the specific use cases. On the other hand, ensuring physical consistency of generated datasets is also crucial. For example, surface pressure could be included to check if geostrophic balance is roughly maintained. This would require more objective and comprehensive evaluation methods to ensure the models' vulnerabilities to extreme windstorm events and their statistical and physical alignments with the input data. Ultimately, improving synthetic data quality and applicability may benefit from a multi-modal framework that combines statistical and application-based evaluation.

## 6. Conclusions

This study explored the application of generative models for simulating windstorms over the UK using ERA5 reanalysis data. The findings show that each model has distinct behaviors. The standard GAN underperformed in representing extremes and maintaining image quality while showing broader variability across multiple runs. The WGAN-GP achieved a more balanced performance in capturing the statistical distributions of typical events but mis-represented extremes occasionally. The U-net diffusion model, while producing smooth and high-quality outputs, consistently underestimated extreme intensities. The diffusion-GAN captures extreme events more effectively and agrees closely with ERA5 in terms of statistical characteristics. However, its occasional overestimation of extreme values again highlights the trade-offs between stability, variability, and the ability to represent extremes in generative modelling.

This study provides a foundation for further research into implementing generative models in meteorology, particularly in windstorm simulations and risk assessment. By supplementing traditional methods with generative outputs, this study could reduce time required to produce large hazardous datasets and potentially enhance decision-making in industries. Future efforts include incorporating temporal dynamics, utilizing multiple datasets to simulate various scenarios, and developing ensemble approaches that combine different models to enhance reliability and diversity.

*Acknowledgements.*

This work is supported by the Department of Meteorology at the University of Reading and the National Centre for Atmospheric Science. YCT acknowledges the guidance received during his master's dissertation, which formed the foundation for this study. Computational resources were provided by the department and the Natural Environment Research Council. KMRH is supported by a NERC Independent Research Fellowship (MITRE; NE/W007924/1).

*Data Availability Statement.*

The ERA5 reanalysis data from the ECMWF are publicly available under the Copernicus Climate Data Store https://cds.climate.copernicus.eu/datasets/reanalysis-era5-single-levels?tab=overview. The code used to train and evaluate our models will be available upon publication.

<div align="center">APPENDIX</div>

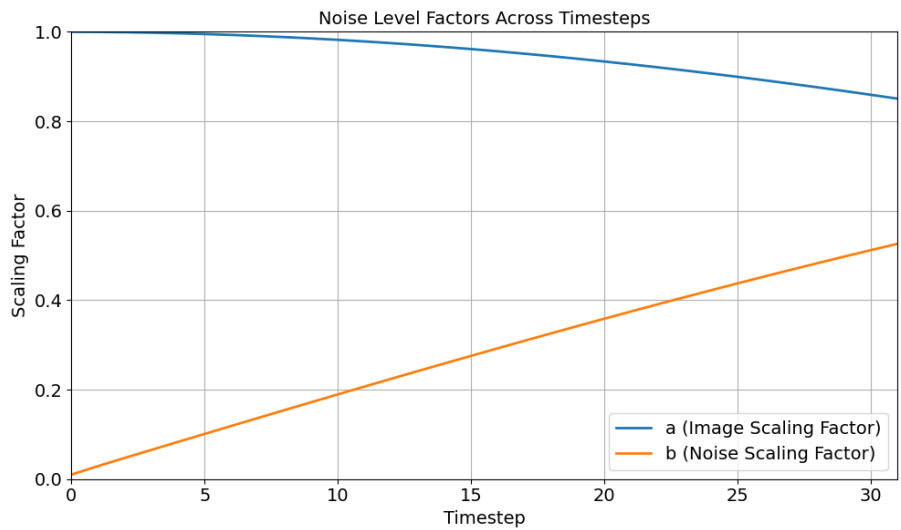

A 1. Noise schedule of the diffusion-GAN model, showing the scaling factors *a* (image factor) and *b* (noise factor) over 32 timesteps.

The following section show some results of failed experiments, which highlights the sensitivity of generative models to configurations and parameter settings. The standard GAN (A 2), without techniques to mitigate instability, exhibits mode collapse (generating nearly identical outputs) and captures variability on the PCA dimensions inaccurately. The WGAN-GP (A 3), despite capturing relatively reasonable spatial patterns and intensities of windstorms, generates outputs with significant impulse noise, potentially caused by a small learning rate and thus poor gradient control. The diffusion-GAN (A 4), with improper noise scheduling, fails to learn and replicate meaningful variations in the data, resulting in pixelated patterns in the outputs. These examples illustrate the challenges of stabilizing training and optimization in generative models.

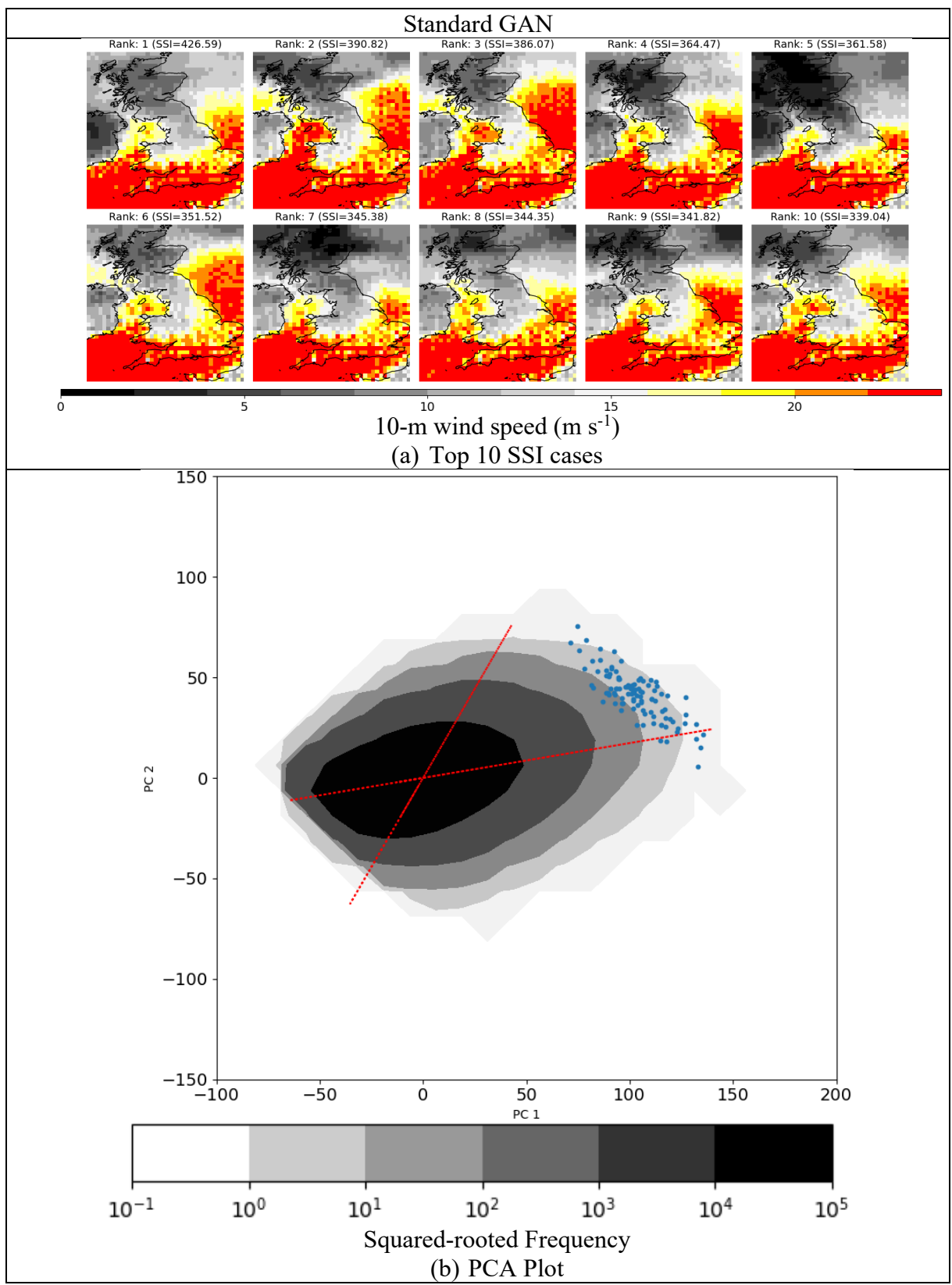

A 2. Top-ranked outputs and PCA plot from the standard GAN without stability techniques.

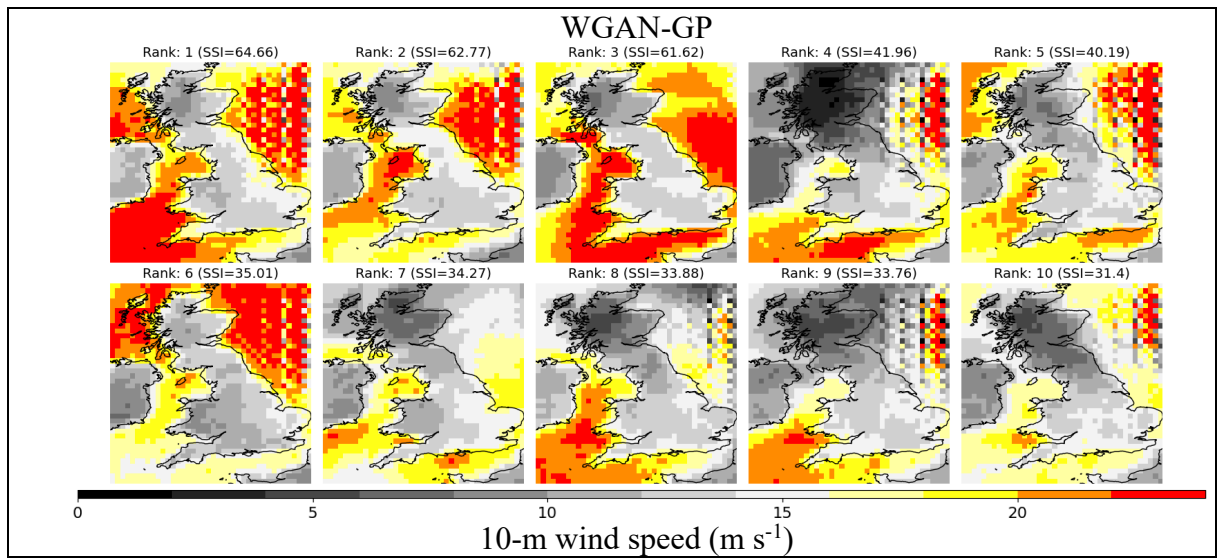

A 3. Top-ranked outputs from the WGAN-GP with a suboptimal learning rate.

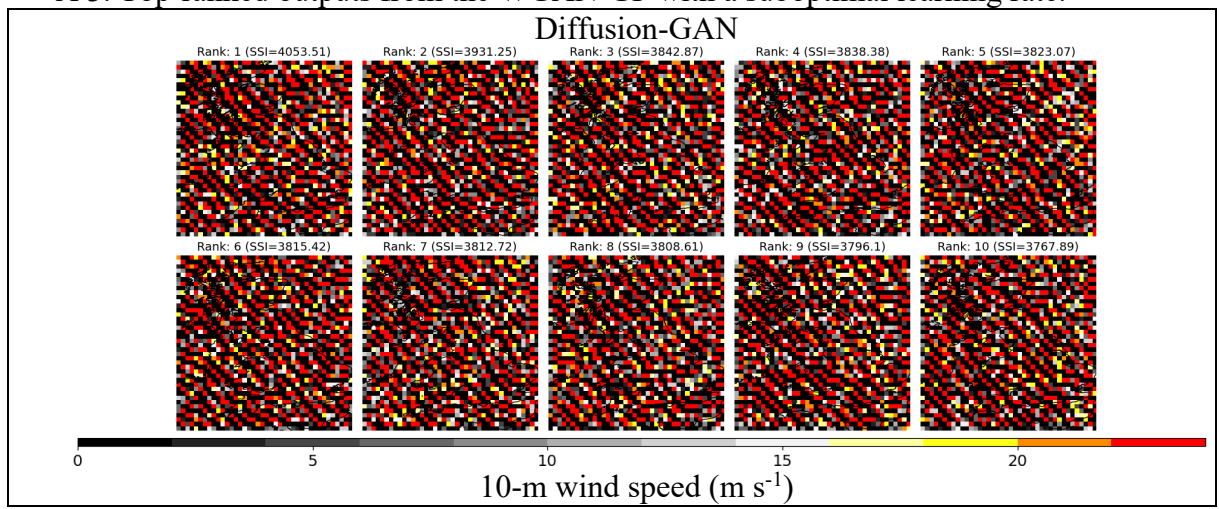

A 4. Top-ranked outputs from the diffusion-GAN with improper noise scheduling.

| Rank | SSI Values | Time | Storm/Event |
|------|-----------|------|-------------|
| 1 | 122.95 | 1400 UTC 25 Jan 1990 | Daria (Burns' Day storm) |
| 2 | 85.37 | 2200 UTC 2 Jan 1976 | Gale of January 1976 (known as Capella in Germany) |
| 3 | 79.63 | 0700 UTC 26 Feb 1990 | Wiebke |
| 4 | 75.17 | 0500 UTC 30 Nov 1954 | |
| 5 | 74.30 | 1300 UTC 16 Sep 1961 | Debbie |
| 6 | 73.19 | 0300 UTC 15 Jan 1968 | Great Glasgow Storm |
| 7 | 72.28 | 0800 UTC 13 Jan 1984 | |
| 8 | 69.60 | 1300 UTC 18 Jan 2007 | Kyrill |
| 9 | 65.49 | 0400 UTC 16 Oct 1987 | Great Storm of 1987 (87J) |
| 10 | 65.15 | 2200 UTC 4 Feb 1957 | |

A 5. Top 10 SSI cases identified from the ERA5 dataset, ranked by their SSI values.

REFERENCES

Adam, E.F., Brown, S., Nicholls, R.J. and Tsimplis, M., 2016: A systematic assessment of maritime disruptions affecting UK ports, coastal areas and surrounding seas from 1950 to 2014. Natural Hazards, 83, pp.691-713.

Ashfaq, M., Rastogi, D., Mei, R., Kao, S.C., Gangrade, S., Naz, B.S. and Touma, D., 2016: High-resolution ensemble projections of near-term regional climate over the continental United States. Journal of Geophysical Research: Atmospheres, 121(17), pp.9943-9963.

Asperti, A., Merizzi, F., Paparella, A., Pedrazzi, G., Angelinelli, M. and Colamonaco, S., 2023: Precipitation nowcasting with generative diffusion models. arXiv preprint arXiv:2308.06733.

Besombes, C., Pannekoucke, O., Lapeyre, C., Sanderson, B. and Thual, O., 2021: Producing realistic climate data with generative adversarial networks. Nonlinear Processes in Geophysics, 28(3), pp.347-370.

Brochet, C., Raynaud, L., Thome, N., Plu, M. and Rambour, C., 2023: Multivariate Emulation of Kilometer-Scale Numerical Weather Predictions with Generative

Adversarial Networks: A Proof of Concept. Artificial Intelligence for the Earth Systems, 2(4), p.230006.

Clark, K.M., 2002: The use of computer modeling in estimating and managing future catastrophe losses. The Geneva Papers on Risk and Insurance. Issues and Practice, 27(2), pp.181-195.

de Melo, C.M., Torralba, A., Guibas, L., DiCarlo, J., Chellappa, R. and Hodgins, J., 2022: Next-generation deep learning based on simulators and synthetic data. Trends in cognitive sciences, 26(2), pp.174-187.

Drozdzal, M., Vorontsov, E., Chartrand, G., Kadoury, S. and Pal, C., 2016: The importance of skip connections in biomedical image segmentation. In International workshop on deep learning in medical image analysis, international workshop on large-scale annotation of biomedical data and expert label synthesis (pp. 179-187). Springer, Cham.

Glorot, X. and Bengio, Y., 2010, March: Understanding the difficulty of training deep feedforward neural networks. In Proceedings of the thirteenth international conference on artificial intelligence and statistics (pp. 249-256). JMLR Workshop and Conference Proceedings.

Goodfellow, I., 2016: Nips 2016 tutorial: Generative adversarial networks. arXiv preprint arXiv:1701.00160.

Goodfellow, I., Pouget-Abadie, J., Mirza, M., Xu, B., Warde-Farley, D., Ozair, S., Courville, A. and Bengio, Y., 2014: Generative adversarial nets. Advances in neural information processing systems, 27.

Gulrajani, I., Ahmed, F., Arjovsky, M., Dumoulin, V. and Courville, A.C., 2017: Improved training of wasserstein gans. Advances in neural information processing systems, 30.

Haylock, M.R., 2011: European extra-tropical storm damage risk from a multi-model ensemble of dynamically-downscaled global climate models. Natural Hazards and Earth System Sciences, 11(10), pp.2847-2857.

Hersbach, H., Bell, B., Berrisford, P., Hirahara, S., Horányi, A., Muñoz-Sabater, J., Nicolas, J., Peubey, C., Radu, R., Schepers, D. and Simmons, A., 2020: The ERA5 global reanalysis. Quarterly Journal of the Royal Meteorological Society, 146(730), pp.1999-2049.

Heusel, M., Ramsauer, H., Unterthiner, T., Nessler, B. and Hochreiter, S., 2017: Gans trained by a two time-scale update rule converge to a local nash equilibrium. Advances in neural information processing systems, 30.

Ho, J., Jain, A. and Abbeel, P., 2020: Denoising diffusion probabilistic models. Advances in neural information processing systems, 33, pp.6840-6851.

Hu, Z. and Hong, L.J., 2013: Kullback-Leibler divergence constrained distributionally robust optimization. Available at Optimization Online, 1(2), p.9.

Ibtehaz, N. and Rahman, M.S., 2020: MultiResUNet: Rethinking the U-Net architecture for multimodal biomedical image segmentation. Neural networks, 121, pp.74-87.

Ioffe, S. and Szegedy, C., 2015: Batch normalization: Accelerating deep network training by reducing internal covariate shift. arXiv preprint arXiv:1502.03167.

Jolicoeur-Martineau, A. and Mitliagkas, I., 2019: Connections between support vector machines, wasserstein distance and gradient-penalty gans. arXiv preprint arXiv:1910.06922, p.9.

Jolliffe, I.T. and Cadima, J., 2016: Principal component analysis: a review and recent developments. Philosophical transactions of the royal society A: Mathematical, Physical and Engineering Sciences, 374(2065), p.20150202.

Klawa, M. and Ulbrich, U., 2003: A model for the estimation of storm losses and the identification of severe winter storms in Germany. Natural Hazards and Earth System Sciences, 3(6), pp.725-732.

Lau, M.M. and Lim, K.H., 2018, December: Review of adaptive activation function in deep neural network. In 2018 IEEE-EMBS Conference on Biomedical Engineering and Sciences (IECBES) (pp. 686-690). IEEE.

Leinonen, J., Nerini, D. and Berne, A., 2020: Stochastic super-resolution for downscaling time-evolving atmospheric fields with a generative adversarial network. IEEE Transactions on Geoscience and Remote Sensing, 59(9), pp.7211-7223.

Minola, L., Zhang, F., Azorin-Molina, C., Pirooz, A.S., Flay, R.G.J., Hersbach, H. and Chen, D., 2020: Near-surface mean and gust wind speeds in ERA5 across Sweden: towards an improved gust parametrization. Climate Dynamics, 55(3), pp.887-907.

Miralles, O., Steinfeld, D., Martius, O. and Davison, A.C., 2022: Downscaling of historical wind fields over Switzerland using generative adversarial networks. Artificial Intelligence for the Earth Systems, 1(4), p.e220018.Ravuri, S., Lenc, K., Willson, M., Kangin, D.,

Potisomporn, P., Adcock, T.A. and Vogel, C.R., 2023: Evaluating ERA5 reanalysis predictions of low wind speed events around the UK. Energy Reports, 10, pp.4781-4790.

Ravuri, S., Lenc, K., Willson, M., Kangin, D., Lam, R., Mirowski, P., Fitzsimons, M., Athanassiadou, M., Kashem, S., Madge, S. and Prudden, R., 2021: Skilful precipitation nowcasting using deep generative models of radar. Nature, 597(7878), pp.672-677.

Rubner, Y., Tomasi, C. and Guibas, L.J., 2000: The earth mover's distance as a metric for image retrieval. International journal of computer vision, 40, pp.99-121.

Santurkar, S., Tsipras, D., Ilyas, A. and Madry, A., 2018: How does batch normalization help optimization?. Advances in neural information processing systems, 31.

Shen, J., Qu, Y., Zhang, W. and Yu, Y., 2018, April: Wasserstein distance guided representation learning for domain adaptation. In Proceedings of the AAAI conference on artificial intelligence (Vol. 32, No. 1).

Szegedy, C., Vanhoucke, V., Ioffe, S., Shlens, J. and Wojna, Z., 2016: Rethinking the inception architecture for computer vision. In Proceedings of the IEEE conference on computer vision and pattern recognition (pp. 2818-2826).

Wang, R., Su, L., Wong, W.K., Lau, A.K. and Fung, J.C., 2023: Skillful radar-based heavy rainfall nowcasting using task-segmented generative adversarial network. IEEE Transactions on Geoscience and Remote Sensing.

Wang, Z., Bovik, A.C., Sheikh, H.R. and Simoncelli, E.P., 2004: Image quality assessment: from error visibility to structural similarity. IEEE transactions on image processing, 13(4), pp.600-612.

Wang, Z., Zheng, H., He, P., Chen, W. and Zhou, M., 2022: Diffusion-gan: Training gans with diffusion. arXiv preprint arXiv:2206.02262.

Wu, Y., Liu, L., Bae, J., Chow, K.H., Iyengar, A., Pu, C., Wei, W., Yu, L. and Zhang, Q., 2019, December: Demystifying learning rate policies for high accuracy training of deep neural networks. In 2019 IEEE International conference on big data (Big Data) (pp. 1971-1980). IEEE.

Yao, W., Zeng, Z., Lian, C. and Tang, H., 2018: Pixel-wise regression using U-Net and its application on pansharpening. Neurocomputing, 312, pp.364-371.